\documentclass[useAMS,usenatbib]{mn2e}
\usepackage{amssymb}
\usepackage{graphicx}
\DeclareGraphicsExtensions{.jpg,.pdf,.png,.eps,.pdf}
\usepackage{url}
\usepackage{color}
\usepackage{tabulary}
\usepackage{multirow}
\usepackage{amsmath}
\usepackage{amsfonts}
\usepackage{ulem}
\usepackage{booktabs, dcolumn}
\usepackage{natbib}

\def\msun{{\rm ~M}_{\odot}}

\def\atg{{\rm ~atg}}

\newcommand\beq{\begin{equation}}
\newcommand\eeq{\end{equation}}
\newcommand{\apj}{ApJ}
\newcommand{\apjl}{ApJ}
\newcommand{\apjs}{ApJS}
\newcommand{\aap}{A$\&$A}
\newcommand{\araa}{ARAA}

\newcommand{\mnras}{MNRAS}
\newcommand{\pasj}{PASJ}
\newcommand{\prd}{PRD}

\newcommand{\nat}{Nature}

\newcommand{\physrep}{Physics Reports}
\newcommand{\bain}{Bull. of the Astr. Inst. of the Netherlands}
\newcommand{\na}{New Astr.}

\begin{document}

\title[BH-BH mergers within the LIGO Horizon]{Binary Black Hole Mergers within the LIGO Horizon:
Statistical Properties and prospects for detecting Electromagnetic Counterparts}

\author[]{Rosalba Perna$^1$, Martyna Chruslinska$^2$, Alessandra Corsi$^3$, 
Krzysztof Belczynski$^4$\\
$^1$ Department of Physics and Astronomy, Stony Brook University, Stony Brook, NY, USA\\
$^2$ Warsaw University Observatory, Al Ujazdowskie 4, 00-478 Warszawa, Poland\\
$^3$ Department of Physics, Texas Tech University, Box 41051, Lubbock, TX 79409-1051, USA\\
$4$ Nicolaus Copernicus Astronomical Centre, Polish Academy of Sciences,
           ul. Bartycka 18, 00-716 Warsaw, Poland}
\maketitle

\begin{abstract}

Binary black holes (BBHs) are one of the endpoints of isolated binary
evolution, and their mergers a leading channel for gravitational wave
events.  Here, using the evolutionary code \textsc{StarTrack}, we
study the statistical properties of the BBH population from
isolated binary evolution for a range of progenitor star metallicities
and BH natal kicks.  We compute the mass function and the distribution of the primary BH
spin $a$ as a result of mass accretion during the binary
evolution, and find that this is not an efficient process to spin up
BHs, producing an increase by at most $a\sim$~0.2--0.3 for very low natal BH
spins.  We further compute the distribution of merger sites
within the host galaxy, after tracking the motion of the binaries in
the potentials of a massive spiral, a massive elliptical, and a dwarf
galaxy. We find that a fraction of 70-90\% of mergers in massive
galaxies and of 40-60\% in dwarfs (range mostly sensitive to the natal
kicks) is expected to occur inside of their hosts.
 The number density distribution at the merger sites further allows us to estimate
the broadband luminosity distribution that BBH mergers would produce,
\textit{if} associated with a kinetic energy release in an outflow,
{which, as a reference, we assume  at the level inferred for the
\textit{Fermi} GBM counterpart to GW150914, with the understanding
that current limits from the O1 and O2 runs would require such
emission to be produced within a jet of angular size within $\lesssim 50^\circ$.}

\end{abstract}

\begin{keywords}
gravitational waves ---  stars: black holes --- gamma
-ray burst: general
\end{keywords}

\section{Introduction}

The LIGO discovery of gravitational waves (GWs) from BBH mergers 
\citep{Abbott2016a,Abbott20162nd,Abbott2017}
has opened a new window onto the universe, providing a definite
confirmation that binary BHs exist, and merge with a local rate
in the range $\sim 12-213$~Gpc$^{-3}$~yr$^{-1}$ \citep{Abbott2016b,Abbott2017}.

The properties of the LIGO-discovered BHs, and in particular the large
masses and relatively low spins, have triggered numerous
investigations aimed at identifying the possible formation channels of
BHs with such properties. Three main mechanisms have been considered:
Classical isolated binary evolution, dynamical
  formation, and chemically homogeneous evolution in tidally
  distorted binary stars.  Within the classical isolated binary
evolution scenario, the binary system undergoes a common envelope (CE)
ejection, or a non-conservative mass transfer
(e.g. \citealt{Tutukov1993, Kalogera2007,Postnov2014}).  Dynamical formation
involves dynamical interactions in dense star clusters
\citep{Sigurdsson1993, Downing2010,Downing2011,Antonini2016,Rodriguez2016, Rodriguez2017}. 
 For example, isolated BHs can acquire a companion via 3-body exchanges and binary-mediated
interactions, which tend to result in the ejections of the lightest
BH.  Finally, for a chemically homogeneous evolution in tidally
distorted binary stars, strong internal mixing occurs as a result of
massive stars being in near contact binaries
\citep{DeMink2009,deMink2016,Mandel2016,Marchant2016}.

During the first (O1) and second (O2) observing runs of advanced LIGO,
a massive follow-up campaign in search for electromagnetic (EM)
counterparts to GWs has been undertaken
(e.g. \citealt{Abbott2016c,Copperwheat2016,Cowperthwaite2016,Evans2016,Kasliwal2016,
  Morokuma2016,Palliyaguru2016,Savchenko2016,
  Smartt2016a,Smartt2016b,Bhalerao2017,Corsi2017,Kawai2017,Racusin2017,Savchenko2017}).
A combined GW-EM detection  would
help break GW parameter degeneracies {(see e.g. \citealt{Pankow2017})}, it would enable measurement of
the source redshift { \citep{Schutz1986,Abbott2017d}}, and help constrain the formation channel of the
BBH binary by probing the environment in which the merger occurred
{ (e.g. dynamical formation \citep{Rodriguez2016} is likely to occur in dense star clusters, while
the classical isolated binary evolution scenario \citep{Kalogera2007} in typical field
galaxies; on the other hand, BBH evolution in an accretion disk around a supermassive black hole
\citep{Bartos2016} will be found in the vicinity of AGNs).} 

To date, most searches for EM counterparts to the LIGO detections have
yielded negative results. Possible exceptions are the potential
$\gamma$-ray counterparts to GW150914 and GW170104 identified by the
\textit{Fermi} (\citealt{Connaughton2016}, but see
\citealt{Greiner2016}) and AGILE \citep{Verrecchia2017} satellites,
respectively.  While in NS-NS and NS-BH mergers EM emission is a
natural outcome of the circularization, and subsequent accretion, of
some mass from the tidally disrupted NS, in the case of BBH mergers
no EM emission is generally expected.  Thus, the tentative
$\gamma$-ray detections by the \textit{Fermi} and AGILE satellites
have received considerable attention, and spurred several ideas {and discussions}
on the possible {presence} of EM counterparts to BBH mergers
\citep{Perna2016, Zhang2016, Loeb2016,Lyutikov2016,Fraschetti2016,
 Woosley2016, Murase2016,Liebling2016,Li2016,Bartos2016,Janiuk2017,Dai2017,
 Stone2017,Ioka2017,DeMink2017,Kimura2017}.

Given the large localization areas of GW events detected by LIGO
\citep{Abbott2016LR}, being able to design an EM follow-up strategy
which optimizes the chances of detection is of paramount importance.
The very first step toward this goal is gaining a better theoretical
understanding of the relation between BBH binary properties, their
merger sites, and the expected brightness of their potential EM
counterparts. However, to date, there exists no specific prediction as
to the most likely sites of BBH mergers within their host galaxies,
and hence as to what afterglow brightness could be expected if these
mergers were to give rise to explosive, gamma-ray burst (GRB)-like
counterparts.

In light of the above, goal of this paper is to make statistical
predictions for the observable properties of BBH mergers, and their
possible EM counterparts, within the specific evolutionary model of
isolated binary evolution, and within the standard GRB afterglow model
\citep[e.g.,][]{Sari1998} for the production of EM counterparts at
various wavelenghts (see also \citealt{Yamazaki2016}).  We build upon previous work of our group
addressed at NS-NS and NS-BH mergers \citep{Perna2002,
  Belczynski2006}.  Using the code \textsc{StarTrack}
\citep{Belczynski2008}, we predict: \textit{(i)} the BBH chirp and
total mass distribution for several combinations of model assumptions
and metallicities; \textit{(ii)} the distribution of merger sites
within their host galaxies; \textit{(iii)} the increase in the spin
distribution of the primary BH due to accretion from the secondary for
several values of the initial spin; \textit{(iv)} the expected
afterglow luminosities in the X-ray, optical, and radio bands. For
this last calculation, we take as a prototype the energetics ($E\sim 10^{49}$~ergs) inferred
for the possible $\gamma$-ray counterpart to GW150914
\citep{Connaughton2016}. While this detection may be spurious
\citep{Greiner2016}, we use it here simply as a potential calibration
point that may need to be changed when (and if) a secure BBH EM
counterpart is detected. Hence, we present our results in a way such
that they can be easily rescaled to different energy values.  { We
  emphasize our agnostic view as to whether EM counterparts to BBH
  mergers can exist. In this light, our main aim for performing this
  statistical computation of EM counterparts is that of providing a
  test for ideas predicting the possibility of an impulsive kinetic energy
  release associated with BBH mergers. More specifically, our
  calculations are meant to provide the link between the ideas
mentioned above (and discussed in more detail in Sec.~4), and the
panchromatic observational campaigns that have (and will) follow
GW-detected BBH mergers. }

The paper is organized as follows: the description of the galactic
potentials used is detailed in Sec.~2, while Sec.~3 describes the
population synthesis calculation, our model assumptions, and the
initial conditions for the simulations.  Sec.~4 presents the
computation of the afterglow spectrum under the assumption that merger
events drive a shock in the interstellar medium.  The results of our
calculations are reported in Sec.~5, and include the distributions for
the chirp and total mass function, the spin distribution of the 
primary BH, and the probability function for the location of merger
events within their host galaxies, for several combinations of the
metallicity and other relevant model parameters.  The predicted
electromagnetic luminosities are provided in three relevant bands.  We
summarize in Sec.~6.

\section{Galaxy potential models}

We consider three galaxy models: a large, Milky Way-type spiral
galaxy, a large elliptical, and a dwarf galaxy. 
The latter is motivated by recent suggestions
\citep{OShau2017} that dwarf galaxies may overabundantly produce
compact binary mergers, and particularly binary BHs.
The type of galaxy is
important for two reasons: \textit{(i)} its potential determines the
motion of the binaries, for given initial speeds and locations;
\textit{(ii)} its medium density (in magnitude and spatial
distribution) influences the brightness of potential EM counterparts
(afterglows).  We follow the motion of the BBH binaries in the model
galaxies from their formation time until the merger (for those which
merge within a Hubble time), taking into account the kick velocities
gained at the formation of each compact object.

The model spiral galaxy consists of a disc and a bulge described by
the \citet{Miyamoto1975} type potential:
\begin{equation}
 \Phi_{MN} = -\frac{G M}{ \sqrt{x^{2}+y^{2} + (a + \sqrt{z^{2} + b^{2}})^{2}} }
\end{equation}
and a dark matter halo described by the \citet{Paczynski1990} potential, 
\begin{equation}
 \Phi_{halo} = \left\{ \begin{array}{ll}
&- \frac{GM}{2 R_{core}} \left[ \log \left( 1  + \frac{r^{2}}{R_{core}^{2}} \right)\right. \\
&+ \frac{R_{core}}{r} \atg \left.\left(\frac{r}{R_{core}}\right)\right] \;\;\;\;\;\;\;\;\;\;\;
    r < R_{cut} \\
&-\frac{GM}{r} + \frac{GM}{R_{cut}} - \frac{G M}{2 R_{core}} \left[ \log
   \left(1+ \frac{R_{cut}^{2}}{R_{core}^{2}}\right)\right. \\
&+ \left.\frac{R_{core}}{R_{cut}} \atg\left(\frac{R_{cut}}{R_{core}}\right) \right]\;\;\;\;\;\;\;\;\;\;\; 
   r > R_{cut}\,. \\
\end{array}\right.
\label{PaczynskiHalo}
\end{equation}
Our model elliptical galaxy consists of a bulge described by the \citep{Hernquist1990} type potential:
\begin{equation}
 \Phi_{H} = -\frac{GM_{E}}{r+a_{E}}
\end{equation}
and a halo with potential given by Eq.~(\ref{PaczynskiHalo}).  

The parameters of the spiral galaxy potential are taken to be the same
as those of the Milky Way: for the disc, $\rm a = 4.2 \ kpc$, $\rm b =
0.198 \ kpc$, $\rm M = 8.78 \times 10^{10} \msun$; for the bulge, $\rm
a = 0 $, $\rm b = 0.277 \ kpc$, $\rm M = 1.2 \times 10^{10} \msun$; and for
the halo, $\rm M_{halo} = 5 \times 10^{10} \msun$, $\rm R_{core} = 6
\ kpc$, $\rm R_{cut} = 100 \ kpc$.
For the elliptical galaxy we assume a mass of $\rm M_{E}=5 \times
10^{11}~\msun$ and a scale factor of $\rm a_{E} = 5$~kpc, while the halo
parameters are assumed to be the same as those for the spiral galaxy.
The fraction of mass in gas is assumed to be $f_{\rm gas}=0.5$ for the bulge and disk, 
and $f_{\rm gas}=\Omega_b/\Omega=0.04$ for the halo \citep{Bahcall+99}.

For the case of the spiral galaxy, binaries are initially placed on
circular orbits in the disc.  We adopt the distribution of stars
within the disc of a spiral galaxy after \citet{Paczynski1990}: 
\begin{equation}
P(R,z) \ dR dz= P(R)dR \ p(z)dz
\end{equation} where:
\begin{eqnarray}
 P(R)dR = a_{R} \ e^{-R/R_{exp}} \frac{R}{R_{exp}^{2}} \ dR\\
R=\sqrt{x^{2}+y^{2}}~~~~\\
a_{R} = \left[ 1 - e^{-R_{max/R_{exp}}} \left(1+\frac{R_{max}}{R_{exp}}\right) \right]^{-1}
\end{eqnarray}
and
\begin{equation}
 p(z)dz = e^{-z/z_{exp}} \frac{1}{z_{exp}} \ dz\,,
\end{equation}
with $\rm R_{exp} = 4.5$~kpc, $\rm R_{max} = 20$~kpc, and $\rm z_{exp} = 75$~pc.

In the case of the elliptical galaxy,  binaries are
placed in the galactic bulge with random orientation of the orbital
angular momentum, and with a mass density corresponding to the
Hernquist potential, $\rho(r)= (M_{E} / 2\pi) a_{E} r^{-1}
(a_{E}+r)^{-3}$.

To model the dwarf galaxy, we take as a proxy a
small spiral galaxy, of total mass (disk + halo + bulge) $M_{\rm
  gal}=1.5\times 10^9~M_\odot$.  Each individual mass component is
rescaled to 0.1\% of the corresponding one for the large spiral, while
the spatial scale parameters of the potential are correspondingly
rescaled to 10\%.

\section{BBH Population Synthesis Models}

To obtain the population of double BH binaries, we use the
\textsc{StarTrack} population synthesis code, described in detail in
\citet{Belczynski2002, Belczynski2008}, with the updated treatment of
the CE phase \citep[as described in ][]{Dominik12},
and the wind mass loss \citep{Belczynski2010}.\\

{Note that in StarTrack we only account for BBH binaries formed within the 
classical isolated binary evolution scenario involving mass transfer 
episodes and/or CE evolution,  as opposed to other
 possible formation channels, such as dynamical formation in dense stellar 
 environments (e.g.  \citealt{Rodriguez2016}), or chemically homogeneous 
 evolution (e.g. \citealt{deMink2016, Marchant2016}).}

\subsection{Initial Conditions and Model Parameters}
For binaries with periods $< 3000 \rm \ days$, we adopt the initial
conditions as described in \citet{Sana2012}, based on spectroscopic
observations of Galactic O-type stars:  Kroupa-like IMF
\citep[IMF][]{Kroupa1993} with the power-law exponent modified to
$-2.3$ for the stars of mass $\rm M> 1.0 \, M_{\odot}$, eccentricity
distribution $f(e) \propto e^{-0.42}$ in the range $[0.0 \, , \,
  1.0]$, period distribution $f(P) \propto \rm [log(P)/{\rm days}]^{-0.55}$ in the
range $[0.15 \, , \, 3.48]$ and a flat binary mass ratio  distribution
$q = \rm M_{\rm b} / \rm M_{\rm a}$, within [0.0, 1.0].  Wider systems cannot be
reliably probed through spectroscopic observations \citep{Sana2012}.
Here, based on the recent review of \citet{Duchene2013}, for systems
with initial periods of over 3000 days, we adopt a mass ratio
distribution $f(q) \sim q^{\gamma}$ with $\gamma = -0.5$.  There is no
clear evidence for the period distribution to change considerably with
periods increasing over 3000 days; we thus keep the rest of initial
distributions of \citet{Sana2012} unchanged and extend the period
distribution up to $\log [P/{\rm days}] = 5.5$. 

We study in depth three evolutionary models, chosen to give
the largest spread in natal kicks, as detailed in
Sec.~\ref{subsec:natal}.  For clarity and consistency with other
published works, we keep the notation consistent with the one adopted
by \citet{Belczynski2016}, and hence we use the same names for the
models that we study. More specifically, we will consider models M10
(standard model), M13 and M15. These are the equivalent of models M1,
M3 and M5 studied by \citet{Belczynski16N}, except that they include
the effect of Pair-instability Pulsation Supernovae (PPSN) and
Pair-instability Supernovae (PSN)\footnote{For the results presented
  here, the main difference between the models with and without PPSN/PSN 
  is in the high mass BH range, $M_{\rm BH}\gtrsim 40M_\odot$,
  hence only of relevance for the tail of the mass function of the
  $Z=1\%Z_\odot$ model.}.

The reference model (M10) has the minimum mass of the primary $\rm M_{mina}
= 20 \msun $, and minimum mass of the secondary $\rm M_{minb} = 15
\msun $, while the maximum allowed mass for a main sequence star is
$\rm M_{max} = 150 \msun $.  The other models (M13/M15) have a minimum mass of
the primary equal to $\rm M_{mina} = 5 \msun $, a minimum mass of the
secondary given by $\rm M_{minb} = 3 \msun $, and maximum
allowed mass as in M10 ($\rm M_{max} = 150 \msun $).  
For each model, we run three sets of metallicities: $\rm Z_\odot =
0.02$, $\rm 10\%~Z_\odot$, $\rm 1\%~Z_\odot$ (the contribution to the
merging population is the biggest for small metallicities, see
\citealt{BelczynskiRepetto16}).  Since the dependence of the remnant
masses on metallicity is much stronger than the dependence on the of
center-of-mass kicks, which (as we explain in what follows) vary among
the models M10/M13/M15, we show results at different metallicities only
for the M1 (standard) model. For this model, $2 \times 10^{7}$
isolated binaries from the above mass range were simulated for each
metallicity, starting from zero-age main sequence (ZAMS) stars. 
For models M13 and M15, we simulated $2 \times 10^{7}$ and $2 \times 10^{6}$
binaries at ZAMS, respectively.
{ The number of BBH systems which merge within a Hubble time is, for model M10,
3254 at $Z_\odot$, 205335 at  $\rm 10\%~Z_\odot$ and 519782 at
$\rm 1\%~Z_\odot$, while it is 1406 for M13 and 435 for M15 (both calculated
at $\rm 10\%~Z_\odot$ only).  Our number of simulated systems, ultimately
limited by computational time, is comparable to, or larger than, what adopted
in other similar studies (i.e. \citealt{OShaughnessy2005a, OShaughnessy2005,
Perna2002}). The properties of BBH systems that merge, in particular for our
standard model M10 at  $\rm 10\%~Z_\odot$, appear statistically robust, as we verified
by computing the resulting distributions by including only a subset of systems. }

\subsection{Natal kicks}
\label{subsec:natal}
Natal kicks play a fundamental role in determining the location
of the merger sites within the host galaxies (and hence the 
medium density encountered by a potential GRB-driven shock). Models M10/M13/M15 
make the following different assumptions
for the natal kicks. In model M10, natal kicks gained by the newly formed BH during
the supernova explosion are drawn from a Maxwellian distribution with the
velocity dispersion $\sigma = 265 \rm \, km \, s^{-1}$, as proposed by
\citet{Hobbs2005} based on observations of proper motions of single,
young pulsars, but their values are lowered proportionally to the
amount of material falling back onto the compact object \citep[see
Eq. 16 of ][]{Fryer2012}.  This model also accounts for a Blaauw kick velocity
(a few km/s), gained by the center of mass of the binary due to mass
loss during supernova explosion \citep{Blaauw1961}.  The amount of
fallback matter increases with the BH mass, hence more massive BHs receive
smaller natal kicks.  Moreover, the most massive ones ($\rm M_{BH}
\gtrsim 10 M_{\odot}$ at solar metallicity) form through direct
collapse, with no supernova explosion, and thus receive no birth kick.
On the other hand, in the 'pessimistic' model M13, BHs receive full natal kicks drawn from a
Maxwellian with $\sigma = 265 \rm \, km \, s^{-1}$ (which leads to
birth velocities of the individal components of $\sim$~400~km/s),
but this time the kicks are not decreased due to fallback. Finally, model M15 has intermediate kicks drawn from a
Maxwellian distribution with the lower velocity dispersion of $\sigma
= 130 \rm \, km \, s^{-1}$ (equivalent of the M15 model in
\citealt{Belczynski16N}).

\begin{figure}
\hspace{-0.3in}
\includegraphics[scale=0.44]{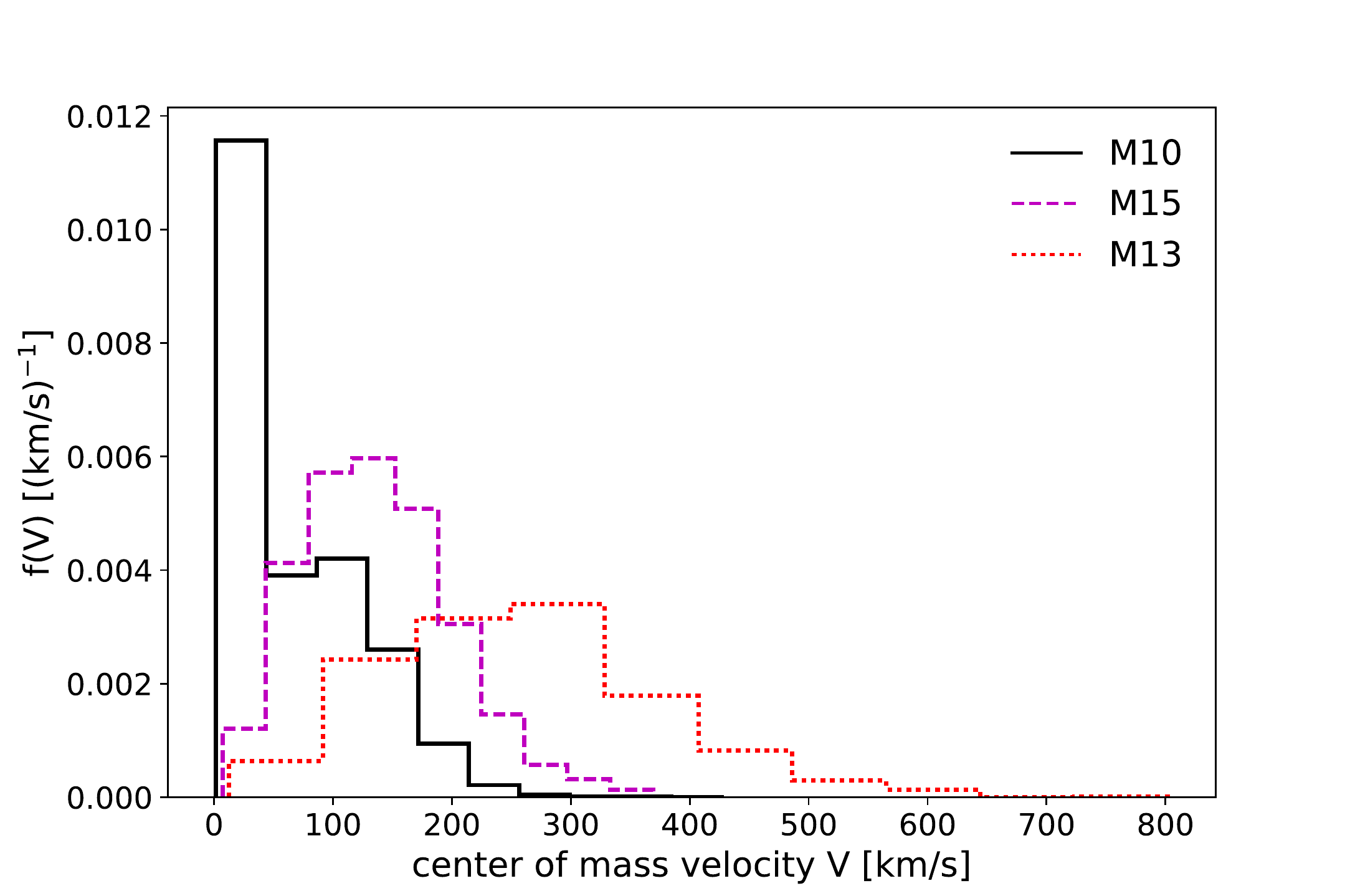}
\caption{The BBH center of mass velocity distribution after the formation
of the second BH, for the three models under consideration.}
\label{fig:vel}
\end{figure}

In order to get a better sense of the role of the natal kicks in the
results for the merger locations within the host galaxies that will be
shown in Sec.\ref{subsec:BinEv}, Fig.~\ref{fig:vel} shows the BBH
center of mass velocity distribution after the formation of the second
BH, for the three models under consideration and $Z=10\% Z_\odot$\footnote{We 
adopt $Z_\odot = 0.02$ \citep{Villante2014}.} (but
the metallicity has very little influence compared to the different
model assumptions discussed above).

\subsection{Binary Evolution}
In the following, we provide a brief outline of our simulations and the main
assumptions behind them. More details
can be found in \citet{Belczynski16N} and references therein.
\begin{itemize}
 \item[(i)] Each binary system is followed from the ZAMS phase, 
and its evolution is assumed to proceed in isolation. 
 \item[(ii)] Binary parameters (mass of the primary, mass ratio, eccentricity,
   separation/period) are randomly drawn  from the distributions
   described in the previous section. 
 \item[(iii)] The binary evolution includes simulations of tides, wind
   mass loss, and interaction between the components via mass transfer.
   These are modeled following the prescriptions detailed in
   \citet{Belczynski2008}.  Depending on the specific combination of
   parameters of each system, the binary may reach the point where the
   first compact object can form, or not (for instance, it can merge
   during the unstable mass transfer).
 \item[(iv)] If there is a supernova, a certain amount of mass is lost
   instantaneously; the magnitude of the natal kick velocity is drawn
   from the assumed distribution (as described in the previous section), with a random
   direction.  The magnitude of the natal kick, the amount of ejecta,
   and the parameters of the binary at the moment of the explosion
   determine whether the binary will be disrupted or not.
     \item[(v)] The evolution of the binary is followed until the formation
       of the second compact object, or until the merger/disruption of
       the binary (if this happens before the formation of the double
       compact object system).
 \item[(vi)] Given the final orbital parameters and the masses of each binary, 
   we can then calculate the merger time due to gravitational radiation. Here we 
   follow only the binaries which  merge within the Hubble time.
\end{itemize}

\begin{figure*}
\includegraphics[scale=0.4]{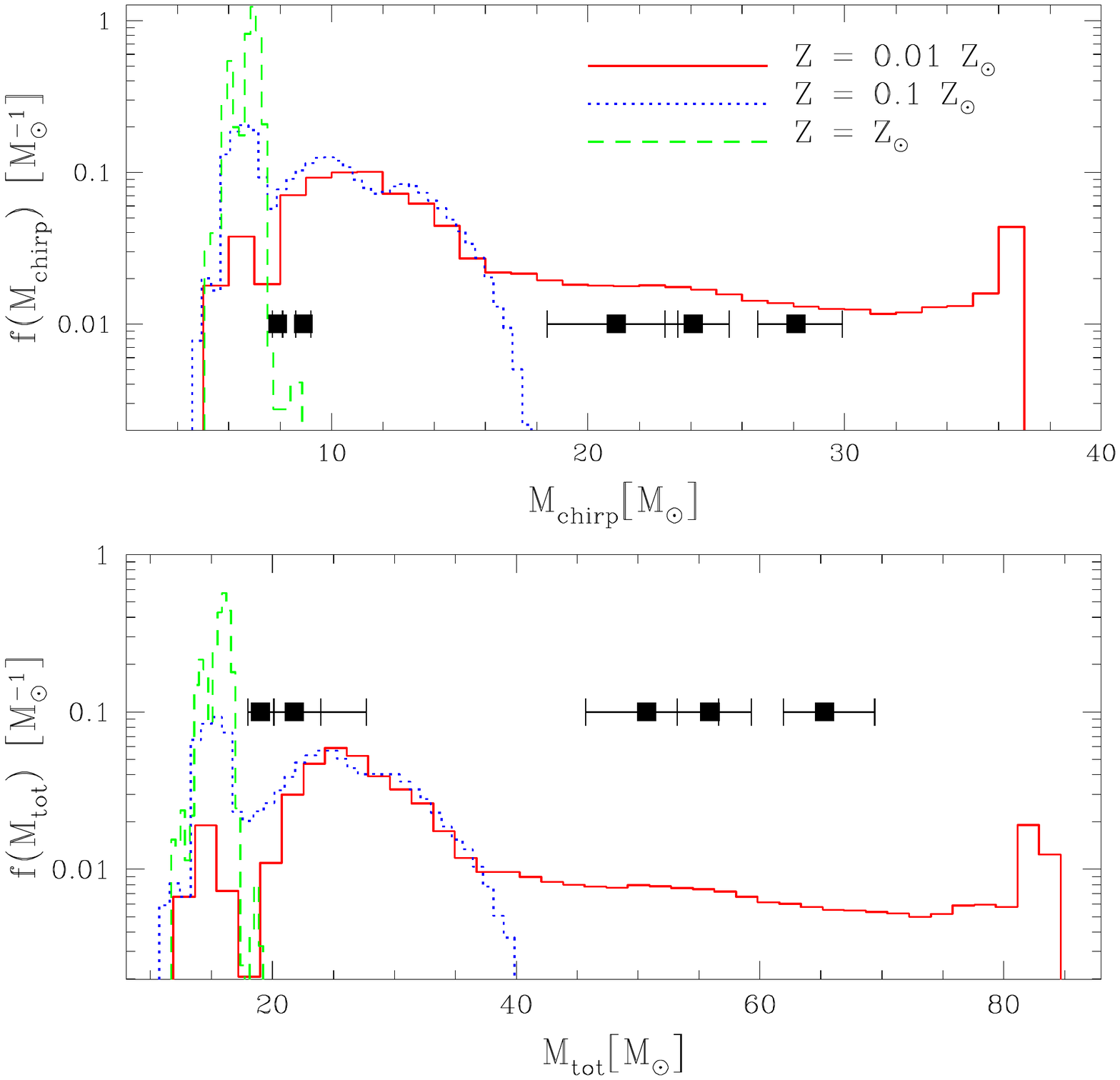}
\includegraphics[scale=0.4]{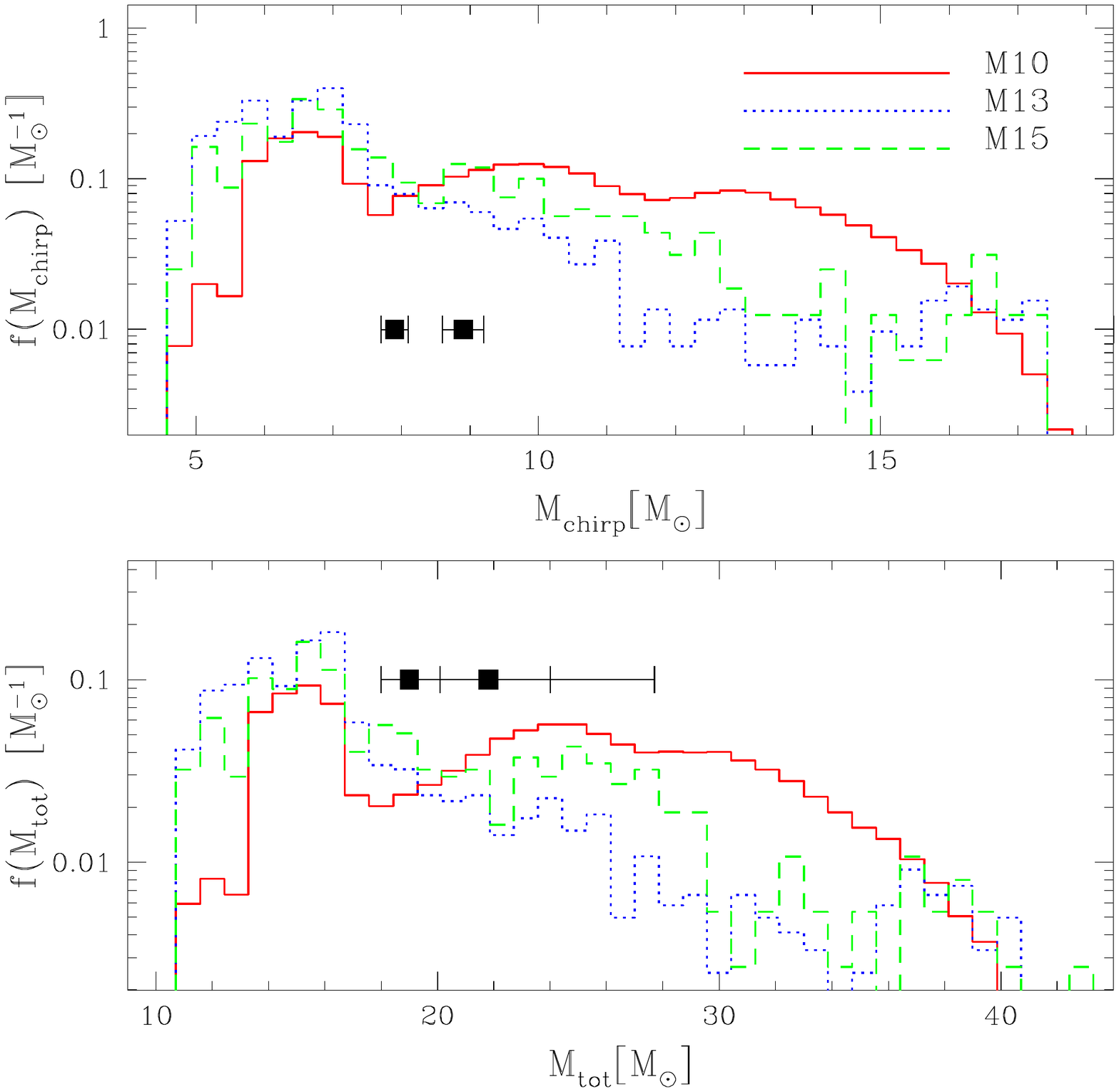}
\vspace{-0.8in}
\caption{\textit{Left:} The distribution of chirp masses (\textit{top panel}) 
and of total masses (\textit{bottom panel})  for the BH
  remnants (in the source frame) for three values of the metallicity,
  in the M10 model. \textit{Right:} Same distributions as in the left panel, but
  for the 3 models studied here, at 10\% of solar metallicity. Note that
  the dependence of the remnant mass fraction on metallicity is much
  stronger than variations in the mass function produced by different
   model assumptions for the natal kicks. {The points with the errorbars
   show the measurements from the first two runs of LIGO (the height at which
   they are placed in the figures has clearly no meaning).}}
\label{fig:chirp}
\end{figure*}

\subsection{BH Spin Evolution due to accretion}
\label{subsec:spin}

{In our simulations BHs can accrete a significant amount of mass during the CE evolution. 
We assume that, during this phase, BHs accrete at 10\% of the Bondi-Hoyle rate (\citealt{Bondi1952}). 
Note that 
this accretion rate is somewhat uncertain, and recent simulations show 
that,  when the density gradients across the CE are taken into account, the rate might be  
lower (e.g. \citealt{MacLeod2015,MacLeod2017}). As a reference with respect to previous related literature,
our total amounts of accreted mass are close to the lower bounds of the calculations by
\citet{OShaughnessy2005}. }

Our formalism for the computation of the change in BH spin as a result 
of BH accretion follows the literature \citep{Shakura1973,Thorne1974}.
In particular, we adopt the formalism of \citet{Brown2000}, also used
by \citet{Belczynski2008}.  

The spin parameter of a BH of mass $M_{\rm BH}$ and angular momentum $J$
is defined as
\begin{equation}
a = {J c \over M_{\rm BH}^2 G}\;,
\label{eq:a}
\end{equation}
where  $G$ is the gravitational constant and $c$ 
the speed of light. If a BH with angular momentum $J_{\rm i}$ accretes an amount of material
with rest mass $M_{\rm acc}$, its new angular momentum will be given
by
\begin{equation}
J=J_{\rm i} + J_{\rm acc}\;,
\label{eq:J}
\end{equation}
\noindent
where
\begin{eqnarray}
J_{\rm acc} &= & \left[{R_{\rm lso}^2 - \tilde{a} \sqrt{2 R_{\rm Sch} R_{\rm lso}} + \tilde{a}^2 
     \over  R_{\rm lso} (R_{\rm lso}^2-{3 \over 2} R_{\rm Sch} R_{\rm lso} + \tilde{a}
     \sqrt{2 R_{\rm Sch} R_{\rm lso}})^{1/2}} \right] \nonumber \\
   & \times & c M_{\rm acc} \sqrt{{R_{\rm Sch} R_{\rm lso} \over 2}} \;. 
\label{eq:Jacc}
\end{eqnarray}
\noindent
Here $R_{\rm Sch}=2GM_{\rm BH}/c^2$ is the BH Schwarzschild radius,
$\tilde{a}\equiv J_{\rm i}/M_{\rm BH}c = a_{\rm i}(GM_{\rm BH}/c^2)$,
with $a_{\rm i}$ the initial BH spin; $R_{\rm lso}$ is the radius of the last
stable orbit
\noindent
\begin{equation}
R_{\rm lso}= {R_{\rm Sch} \over 2} \{3+r_2-[(3-r_1)(3+r_1+2 r_2)]^{1/2}\} 
\label{eq:Rlso}
\end{equation}
with
\noindent
\begin{equation}
r_1= 1+\left(1- {4 \tilde{a}^2 \over R_{\rm Sch}^2}\right)^{1/3} 
\left[ \left(1 + {2\tilde{a} \over R_{\rm Sch}} \right)^{1/3} + 
\left(1 - {2\tilde{a} \over R_{\rm Sch}} \right)^{1/3} \right] \nonumber \\
\label{eq:r1}
\end{equation}
and 
\noindent
\begin{equation}
r_2= \left( 3 {4 \tilde{a}^2 \over  R_{\rm Sch}^2} + r_1^2  \right)^{1/2}.   
\label{eq:r2}
\end{equation}
Note that, as the BH accretes mass, its gravitational mass increases as
\begin{equation}
M_{\rm BH,f} = M_{\rm BH}\;+\;\frac{E}{c^2}\;,
\label{eq:Mf}
\end{equation} 
where
\begin{equation}
E= M_{\rm acc}c^2 \left[\frac{R^2_{\rm lso}-R_{\rm Sch}R_{\rm lso} + \tilde{a}
\sqrt{R_{\rm Sch}R_{\rm lso}/2}}{R_{\rm lso}(R_{\rm lso}^2-
\frac{3}{2} R_{\rm Sch}R_{\rm lso} + \tilde{a}\sqrt{2R_{\rm Sch}R_{\rm lso}})^{1/2}}
 \right]\;.
\label{eq:energy}
\end{equation}

\section{Possible EM counterparts to BBH merger events}
\label{subsec:model}

\subsection{Theoretical arguments and current observational constraints}

Whether BBH mergers can be accompanied by a
release of energy in the form of a relativistic outflow, is a
completely open question at the moment. {This is true from both
  the observational and theoretical points of view. Observationally,
  the support comes from the tentative detection of a short
  $\gamma$-ray counterpart to GW150914 by \textit{Fermi} GBM
  \citep{Connaughton2016}, and to GW170104 by AGILE
  \citep{Verrecchia2017}. The significance of the
former detection has been challenged by \citet{Greiner2016},
but more recently \citet{Connaughton2018} has argued against
the criticism and reaffirmed their earlier conclusions. 
Independent analysis of the GBM data by another team \citep{Bagoly2016}
also found a potential counterpart near the time of GW150914.

Assuming that the detection was indeed real, then the question arises
as to why none of the other BBH events in the LIGO runs O1 and O2 has
had any detection, and what that implies in terms of constraints on
the possible $\gamma$-ray energetics of these events.  This question
is currently the subject of an in-depth study by the \textit{Fermi}
team (V. Connaughton, private communication); however, a
back-of-the-envelope estimate can be made by simply
using the distances to the BBH GW sources, and the \textit{Fermi} 
sensitivity to short bursts, to derive the minimum observable luminosity
 for each of the sources.
Adopting as a reference detection sensitivity 
to short bursts the least fluent burst detected by \textit{Fermi} \citep{Bhat2016},
$F_{\rm min}= 2.2\times 10^{-8}$~erg~cm$^{-2}$~s$^{-1}$, we derive a
minimum detectable luminosity of $L_{\rm min}\sim2.6\times 10^{46}$~erg~s$^{-1}$~$D^2_{100}$, where
$D_{100}$ is the distance to the source in units of 100~Mpc. With the distances
to the BBH GW events ranging between 340~Mpc for GW170608 to 880~Mpc
for GW170104, we derive luminosity limits between $\sim 3\times 10^{47}$~erg~s$^{-1}$ 
and $\sim 2\times 10^{48}$~erg~s$^{-1}$. 
These upper limits on the luminosities are lower than what 
inferred for the potential \textit{Fermi} counterpart to GW150914 ($L\sim 10^{49}$~erg~s$^{-1}$).
In order to assess their significance in constraining emission
models, we need to estimate what is the actual detection probability of an event with the 
\textit{Fermi} GBM. 
To this aim, we need to account for the fact that the GBM had about 60
- 80\% coverage of the large LIGO annuli for the BBH mergers for which
they had data, and that about 15\% of the time they are turned off in
the South Atlantic Anomaly. All together, this yields a roughly 50\%
chance of directly observing anything that happens anywhere in the sky
(V. Connaughton, private communication).  Considering a total of 6
events (including also the LIGO/Virgo trigger LVT151012 to be more
conservative), the above estimates would then imply that about 3 out
of the 6 events should have been detected by the \textit{Fermi} GBM
\textit{if the $\gamma$-ray emission was isotropic.} A tentative detection of
$\lesssim 1$ events would then imply that any putative $\gamma$-ray
radiation would have to be produced within a jet of radial angular size $\theta_{\rm jet}$ 
such that its beaming factor, $(1-\cot\theta_{\rm jet})$, is $\lesssim 1/3$. This yields  $\theta_{\rm jet}
\lesssim 50^\circ$ (assuming two jets of the same size). 
To put this limit within the context of observations, the closest phenomenology
we can rely on is that of the standard short GRBs. For these, a comprehensive
analysis over a decade of data \citep{Fong2015} showed that the median opening
jet angle is $\sim16^\circ$, which would be consistent with the current limits 
on non-detections of $\gamma$-rays from BBH mergers from geometric considerations alone.
Therefore, the current $\gamma$-ray limits do not pose strong constraints
on the presence of $\gamma$-ray radiation, unless there were reasons to believe
that this had to be much more isotropic than what inferred for the short GRBs\footnote{ 
Note that in the case of binary NS mergers, an additional 
  wide-angle, low-luminosity emission is expected from a cocoon produced
  when the relativistic jet propagates within the merger ejecta
  (e.g. \citealt{Lazzati2017}), but in the case of BBH mergers there
  would be no such ejecta for the jet to interact with.}.}

{From a theoretical point of view, the tentative $\gamma$-ray
  counterparts discussed above have led to a} number of ideas being proposed
for producing impulsive energy releases in association with a BBH
merger.  These include, among others {(an extensive list can be found in 
the introduction)}, accretion from the envelope of a
star whose core fragmented and collapsed into the two merging BHs
(\citealt{Loeb2016}, but see \citealt{Woosley2016}), 
or a wind driven by a Poynting flux from charged BHs
\citep{Zhang2016}, or accretion from a fallback disk around one of the
BHs \citep{Perna2016}. In this last model, the disk can survive for a
long time as a dead disk, due to cooling and the suppression of
viscosity once the magnetorotational instability is shut down; the
disk is then revived and rapidly accretes during the final inspiral
phase of the BBH merger. {Note that a total energy reservoir of
$\sim 10^{49}$~ergs as argued for GW150914 is something that could
be easily produced in this astrophysical scenario. In fact, the
stellar model used by Perna et al. (2016)  was taken from a suite of standard pre-supernova
profiles computed with the MESA code \citep{Paxton2013} in an earlier
work \citep{Perna2014}, and hence not fine-tuned for this particular event. }

{With severeal ideas in the literature for an energy reservoir
  connected to a BBH merger, the next question is whether a
  relativistic outflow can be driven.  This issue has a long history
  of being investigated within the context of binary neutron stars
  mergers, where Magnetohydrodynamical (MHD) simulations in full General
  Relativity (GR) have given hints of a magnetically-driven outflow
  \citep{Rezzolla2011,Ruiz2016,Kawamura2016,Ciolfi2017}. 
Very recently, and of direct interest to this work, \citet{Khan2018}
performed GRMHD simulations of disk accretion onto BH binaries,
and observed  collimated and magnetically-dominated
outflows emerge in the disk funnel independently of the disk
model. These important simulations lend support to the idea that, if there
is a minidisk at the time of the BBH merger, then an outflow-induced shock
can be driven into the medium, and generate an afterglow-like radiation
similar to what observed in the case of the binary neutron star merger
\citep{Abbott2017c}. Additionally, as pointed out by \citet{Murase2016},
if a minidisk around the merging BHs were to exist, its thermal emission
could be seen as an additional optical transient, lasting from hours to days. 
Shock heating of material surrounding the BBH site is also likely to produce 
broad-band radiation, from medium-energy X-rays to infrared \citep{DeMink2017}. 
These authors further discuss various mechanisms (in addition to the
supernova fallback discussed above) by which the progenitor
stars of BHs can shed mass during their evolution, including common-envelope
ejection, eruptive mass-loss episodes, centrifugally shedding after spin up, 
mass shedding during non-conservative Roche-lobe overflow. The amount
of mass available via these mechanisms can be of several solar masses, and
hence easily able to accomodate the requirement of $\sim 10^{-4}M_\odot$
set by the \textit{Fermi} counterpart (see also \citet{Janiuk2017} for another
binary evolutionary model leading to remnant mass at the time of the BBH merger). 
 An abundant gas reservoir to tap
energy from can also be of a different origin than the progenitor stars
if the BBH is trapped in the inner region of the accretion disk of an Active Galactic Nuclei,
in which case the BBH can accrete a significant amount of mass from the disk
\citep{Bartos2016,Stone2017}.

Last, as an alternative to an energy reservoir due to matter as in the
models discussed above, the BBH power could also be of electromagnetic
nature due to a residual charge in the merging BHs
\citep{Zhang2016,Liebling2016}. The simulations by
\citet{Liebling2016} showed that a BBH such as the one observed in
GW150914 can satisfy the energy requirements of the \textit{Fermi} GBM
counterpart with a charge of $Q/M = 10^{-4}$ assuming good radiative
efficiency.  }
  
Whether any of these scenarios (or others) is realized in nature when
two BHs merge is a question of great interest in astrophysics. For
this reason, each LIGO detection of a BBH merger event has been
followed by a massive observational campaign in a broad wavelength
range. EM follow ups are expected to continue in future rounds of
LIGO/Virgo runs.

The calculations in this paper reported so far allow us to proceed
further and make an estimate {of the broad-band radiation expected
in association to high-velocity outflows propagating in the interstellar medium. }
This is because a  shock would be a generic
feature \citep{Cavallo1978},
rather independent of the specific details of the mechanism that
produced the energy release. As the blastwave plows into the interstellar medium and
slows down, it emits radiation over a broad-band energy range, the
so-called afterglow.  In the following (cfr. Sec.5.3), by taking a fully agnostic
approach as to whether a (non-GW) energy release accompanies a BBH
merger, we aim at providing predictions for this afterglow at several
bands of observational interest, with the goal of providing a
theoretical framework against which future observations can put to
test a range of theoretical models such as the ones discussed
above. {Our population-synthesis simulations yield the 
density distribution expected from a population of field BBHs, and hence
this is a robust aspect of the calculation. For the energetics, we
adopt as a reference the value required by the \textit{Fermi} counterpart,
since this has been the one mostly used in the literature, and the proposed
theory models are generally able to accomodate it\footnote{ However, given the anaytical
scalings of the afterglow luminosity with energy, our results can be easily
rescaled}.} 

Using our estimates of afterglow-like EM counterparts, we then perform a first comparison using
the available data from the past GW events, and hence predict how many
BBH merger events are needed in order to rule out models predicting
an energy release comparable to what required by the \textit{Fermi}
counterpart.  

\subsection{Afterglow model for relativistic outflows}

In the following we adopt the generic afterglow model developed by
\citet{Sari1998} for a relativistic, adiabatic shock.  Radiation is
expected as a result of synchrotron emission by1 shock-accelerated
electrons. We compute such emission within the standard synchrotron
model The electrons are assumed to have a power-law distribution of
Lorentz factors $\gamma$ above a minimum value $\gamma_m$, and are
able to loose rapidly energy to radiation if their $\gamma$ is larger
than a threshold value $\gamma_c$. The resulting spectral energy
distribution is then \citep[e.g.,][]{Sari1998} \beq F_\nu =
F_{\nu,{\rm max}} \left\{
  \begin{array}{ll}
     (\nu/\nu_c)^{1/3}, & \hbox{$ \nu < \nu_c$} \\
     (\nu/\nu_c)^{-1/2}, & \hbox{$ \nu_c\le \nu < \nu_m$} \\
     (\nu_m/\nu_c)^{-1/2}(\nu/\nu_m)^{-p/2}, & \hbox{$ \nu \ge \nu_m$} \\
  \end{array}\right.\;,
\label{eq:Fnu1}
\eeq
where $\nu_c\equiv\nu(\gamma_c)$ and $\nu_m\equiv\nu(\gamma_m)$.

On the other hand, if the criterium $\gamma_c>\gamma_m$ is satisfied,
only electrons with $\gamma_e > \gamma_c$ can cool efficiently.  Under
this condition, the spectrum takes the functional form \citep[e.g.,][]{Sari1998}
\beq 
 F_\nu  = F_{\nu,{\rm max}} \left\{
  \begin{array}{ll}
     (\nu/\nu_m)^{1/3}, & \hbox{$ \nu < \nu_m$} \\
     (\nu/\nu_m)^{-(p-1)/2}, & \hbox{$ \nu_m\le \nu < \nu_c$} \\
     (\nu_c/\nu_m)^{-(p-1)/2}(\nu/\nu_c)^{-p/2}, & \hbox{$ \nu \ge \nu_c$} \\
  \end{array}\right.\;.
\eeq 
In the above equations, the parameter $p$ represents the power-law index of 
the electron energy distribution. The maximum flux intensity is achieved when 
$F_{\nu}=F_{\nu,{\rm max}}$, where \citep{Sari1998}
\beq
F_{\nu,{\rm max}}= 110\;n^{1/2}{\xi_B}^{1/2}\;E_{52}d_{28}^{-2}\;(1+z)\;{\rm mJy}\;.
\label{eq:Fnu2}
\eeq

This formalism assumes that the magnetic field energy density in the
shock rest frame is a fraction $\xi_B$ of the equipartition value, and
the electrons carry a fraction $\xi_e$ of the dissipated energy.
$E_{52}$ is the explosion energy in units of $10^{52}$~erg, while
$d_{28}$ is the luminosity distance in units of $10^{28}$~cm, and
$t_d$ is the time in days, as measured in the observer frame, since
the beginning of the burst. The relevant frequencies at which the
spectral index changes are the cooling frequency $\nu_c$
(corresponding to energies at which radiative losses over the shock'€™s
lifetime are significant) and the synchrotron frequency $\nu_m$ (the
minimal energy of the radiating electrons), respectively given by the
following expressions \citep{Sari1998}
\beq
\nu_c(t) = 2.7\times 10^{12}\;{n_1}^{-1} {\xi_B}^{-3/2} 
E_{52}^{-1/2}t_{\rm d}^{-1/2}\;(1+z)^{-1/2}\;{\rm Hz}\;,
\label{eq:nuc}
\eeq
and \citep{Sari1998}
\beq
\nu_m(t)=5.7\times 10^{14}\;{\xi_e}^2 {\xi_B}^{1/2} 
E_{52}^{1/2}t_{\rm d}^{-3/2}\;(1+z)^{1/2}\;{\rm Hz}\;.
\label{eq:num}
\eeq 
Additionally, at low frequencies, synchrotron self-absorption can become important.
The self-absorption frequency below which the spectrum becomes optically thick
is given by \citep{yost2003}
\beq
\nu_a(t)=4.2\times 10^{8}f(p)\;({\xi_e^*})^{-1} {\xi_B}^{0.2} 
E_{52}^{0.2}\;n^{0.6}\;(1+z)^{-1}\;{\rm Hz}\;,
\label{eq:nua}
\eeq 
where $f(p)=[(p+2)(p-1)/(3p+2)]^{0.6}$ and $\xi_e^*=\xi_e (p-2)/(p-1)$.
Below this frequency, the spectrum declines more rapidly, as $\nu^2$.

\section{Results}
\subsection{Chirp mass, total mass, and merger sites}
\label{subsec:BinEv}
The left panel of Fig.~\ref{fig:chirp} shows the distribution of chirp
mass, $M_{\rm chirp}=M_1^{3/5}M_2^{3/5}(M_1+M_2)^{-1/5}$, and of total mass, 
$M_{\rm tot}=M_1+M_2$, for our three
  representative metallicities in model M10, while the right panel
  shows the same distributions for $Z=10\% Z_\odot$, and our three
  representative models (M10/M13/M15). Note that, for relatively low mass GW events,
  the GW signal in the LIGO sensitive frequency band is dominated by
  the inspiral phase, and the chirp mass is the most accurately
  measured parameter \citep{Ajith2009}.
As expected, the mass distribution has a strong
  metallicity dependence, reflecting the significant amount of mass
  lost to winds during the evolution of the more metal rich progenitors
stars. On the other hand, the dependence on the specific model at
the same metallicity is considerably weaker, with model M10 displaying
an enhancement in mass at the higher end of the distribution. This
results from the larger amount of fallback for the more massive BHs.
{The points overlaid on the plot show the mass measurements from the
first two observational runs \citep{Abbott2016m1,Abbott2017PhRvL,Abbott2017a,Abbott2017-19m}. 
A comparison between theory and data suggests that, within the isolated
binary formation scenario, the BBHs which merged where likely formed within
a range of metallicities. However, since the
distributions  of merger locations are dependent only very weakly on the progenitor stars
metallicity, we can safely adopt a single value of metallicity for the calculations which
mainly depend on locations.  }

 \begin{figure}
 \hspace{-0.3in}
\includegraphics[scale=0.44]{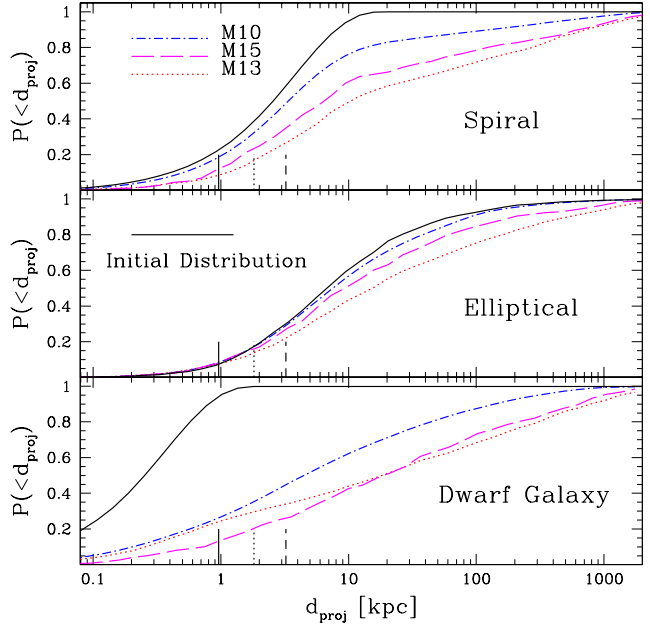}
\vspace{-0.5in}
\caption{The fraction of BBH binaries which merge within a projected
  distance $d_{\rm proj}$ of their host galaxy center. Distances are
  the largest in the M13 model, reflecting the larger kicks at birth. A
  fraction $\sim 10-30\%$ (with precise value dependent on the model)
  is expected to be found outside of their elliptical / spiral
  hosts (formal size set to 100~kpc). The fraction of BBH merger events expected to occur
    outside of dwarf hosts (formal size set to 10~kpc) is larger ($\sim 40-60\%$) due to the
    shallower galaxy potential.  For each
  model, the dependence of distances on metallicity is negligible,
  hence only the $Z=0.1Z_\odot$ case is shown.  The three
    vertical lines correspond to an angular distance of 1\,arcsec
     for a galaxy at a redshift of $z=0.05$ (solid line), $z=0.1$
  (dotted line) and $z=0.2$ (dashed line).}
\label{fig:dist}
\end{figure}

\begin{figure}
\hspace{-0.3in}
\includegraphics[scale=0.44]{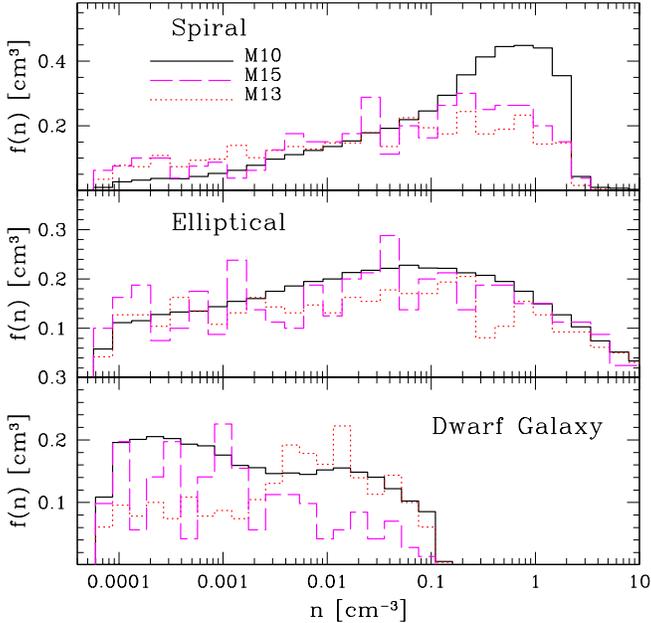}
\vspace{-0.5in}
\caption{The density distribution at the sites of BBH mergers, for
our galaxy potentials and three evolutionary models.  Densities
tend to be higher in the M10 model, reflecting the smaller natal
kicks of the BHs. The densities around the merger sites are important
for the possible production of a shock-driven afterglow.}
\label{fig:dens}
\end{figure}

In Fig.\,\ref{fig:dist} we show the distribution of projected
distances (from the galaxy center) of the merger sites for our three
representative models and the three galaxy types, for $Z=0.1Z_\odot$.   The orientation
angle of the galaxy with respect to the line of sight in each Monte
Carlo realization is drawn from a random distribution. All three
models predict a sizable fraction of mergers to occur at very large
distances, with a fraction $\sim 10-30\%$ (depending on the model)
expected to occur outside of spiral and elliptical galaxies (formal
size set at 100~kpc). As expected, model M13, with the largest natal
kicks for the BHs, is the one which predicts the largest fraction of
mergers to be located at the largest distances.  The fraction of BBH
merger events expected to occur outside of dwarf hosts (formal size
set at 10~kpc) is larger, ranging from $\sim 40\%$ in the model with
the smallest natal kicks to $\sim 60\%$ in the model with the largest
kicks. This larger fraction is expected as a result of the shallower
potential of the dwarf galaxy.

As a reference for the observability of a putative EM counterpart, the
figure also shows, with vertical lines, the projected distances
corresponding to 1~arcsec for a galaxy at the three representative
redshifts of $z=0.05, 0.1, 0.2$, all within the aLIGO horizon, at
least for the most massive BHs. This shows that observational
facilities with sub-arsec localization capabilities could measure
potential counterpart off-sets / positions within the host galaxy.

 \begin{figure*}
\centering
\includegraphics[width=5.81cm]{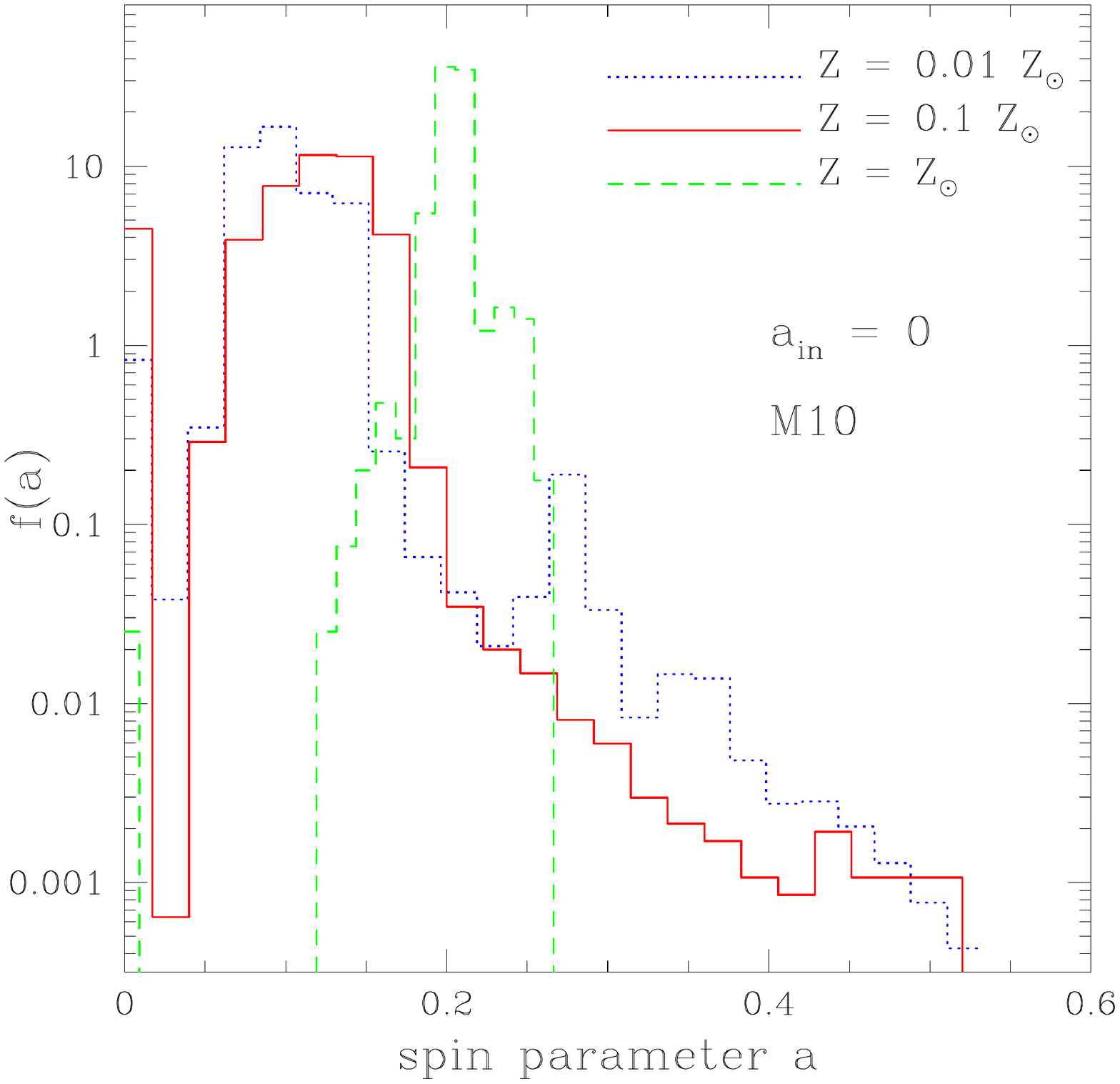}
\includegraphics[width=5.81cm]{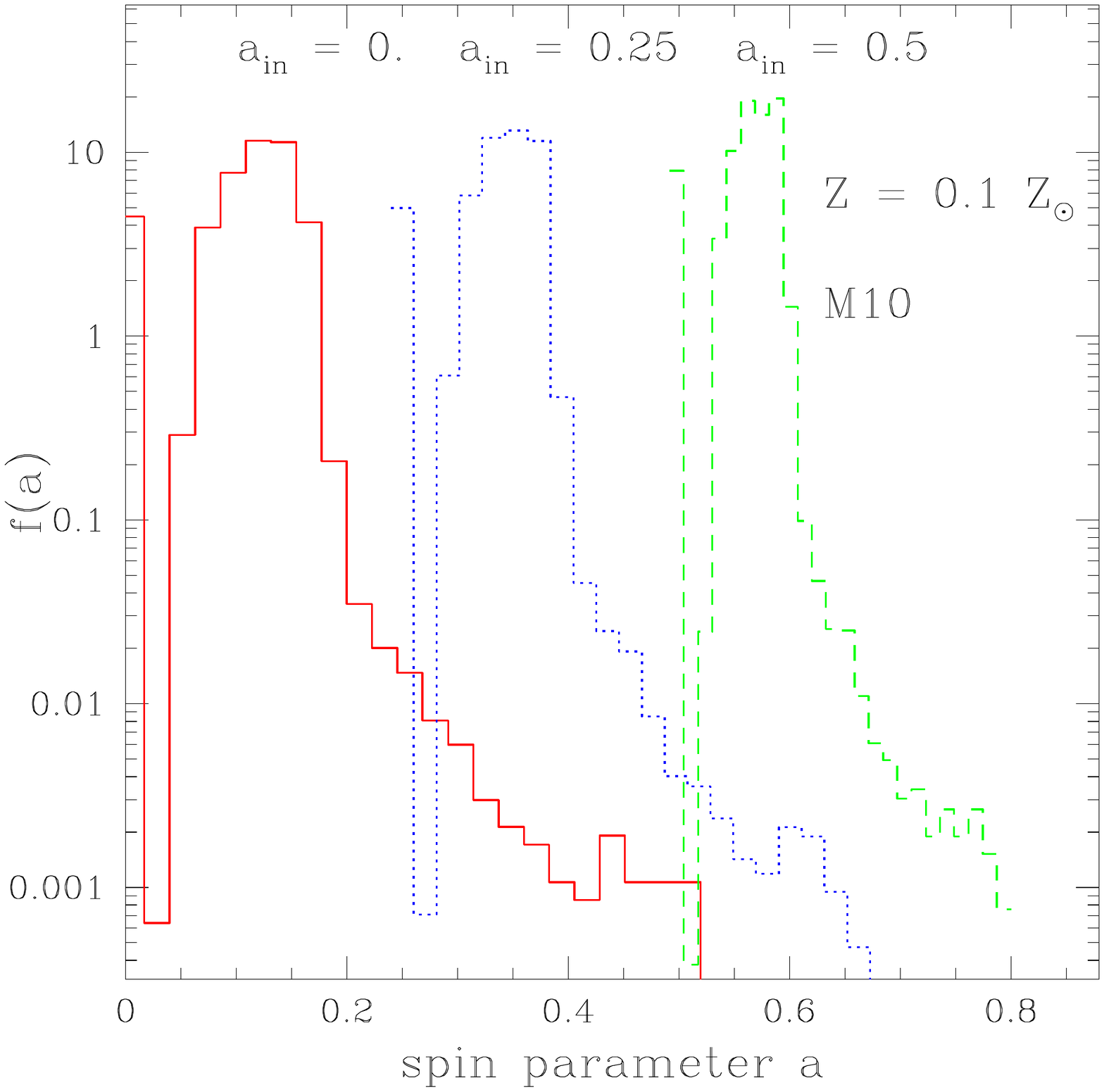}
\includegraphics[width=5.81cm]{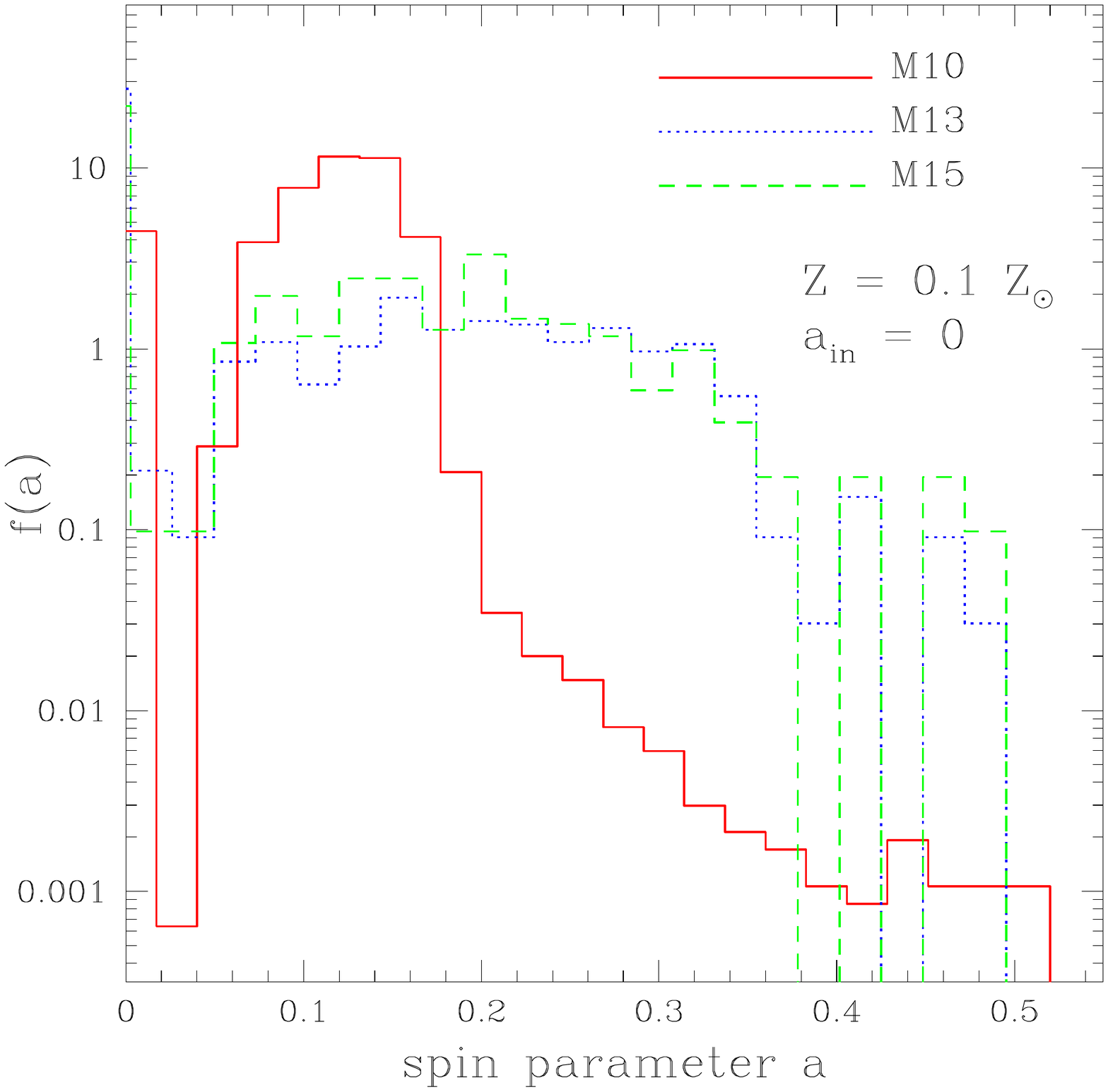}
\vspace{-0.5in}
\caption{The distribution of the spin parameter of the primary BH due
  to accretion of matter from the companion during the binary
  evolution. \textit{Left:} Dependence on metallicity for the M10
  model, assuming that the initial spin of the BH (following the SN
  explosion) is zero. \textit{Middle:} Dependence of the spin distribution
on the initial BH spin, for the M10 model at 10\% of solar metallicity. \textit{Right:}
The spin distribution in the three evolutionary models studied here, at 10\% metallicity
and assuming zero initial spin.}
\label{fig:spin}
\end{figure*}

For the computation of afterglow-like EM radiation accompanying BBH
mergers, the ambient density of the medium at the merger sites
plays an important role, since the afterglow peak flux scales as
$n^{1/2}$ (see Eq.~\ref{eq:Fnu2}).  Fig.\, \ref{fig:dens} shows the number
density distribution predicted by our models, for evolution in a
spiral, elliptical, and dwarf galaxy. Higher ambient densities are
generally expected in model M10, given the smaller distances that the
binaries have traveled prior to merging. Correspondingly, this
scenario will predict brighter afterglows, as quantified
in Sec.\,\ref{subsec:results}.

\subsection{Spin evolution of the primary BH as a result of accretion during binary evolution }

In the binary evolutionary scenario under consideration, the
calculation of the spins that we perform -- following the formalism
described in Sec.~\ref{subsec:spin} -- and report on here refers to
the evolution of the spin of the primary BH (i.e. the BH born first,
which is typically the more massive of the two BHs at merger time) as
a result of mass accretion from the companion star during evolution.
As evident by Eq.~(\ref{eq:J}), the angular momentum (and hence the
spin) of the BH following accretion depends on the amount of angular
momentum possessed by the accreted mass, as well as on the initial
angular momentum possessed by the BH immediately following the
supernova (SN) explosion. Since the distribution of BH spins following
the SN explosion is not well known, here we perform calculations for
three initial values of the primary BH spin (namely $a_{\rm in}=0,
0.25, 0.5$), so that the net effect of the increase due to accretion
(which we accurately compute here) can be immediately inferred.

The left panel of Fig.~\ref{fig:spin} shows the expected spin
distribution in model M10 for the three representative metallicities
and an inital BH spin equal to zero. In all the cases, the spin gained
by the BHs is rather marginal. The distribution peaks at $a\sim 0.1$,
for the 1\% and 10\% solar metallicities, and at $a\sim 0.2$ for the
solar metallicity case. The larger spin acquired by the BHs from
higher metallicity stars is due to the fact that, while the accreted
mass during the common evolution phase is not significantly dependent
on metallicity, however, the mass of the BHs which are accreting is
typically smaller at higher metallicities, hence resulting in a larger
spin.

The middle panel of Fig.~\ref{fig:spin} shows the final spin
distribution for model M10 at 10\% metallicity, for three initial values
of the initial spin of the BHs at birth. The general trend is that of a
relatively smaller increase for the largest initial spins, as expected. 
Last, the right panel shows the
dependence of the spin distribution on the different evolutionary model
assumptions. Compared to model M10, models M13 and M15 display a longer
tail of larger spins. This is mainly due to the fact that model M10 produces
a relatively higher fraction of more massive BHs compared to models M13 and M15.

So far, LIGO measurements of BH spin magnitudes have been inconclusive \citep{Abbott2017PhRvL}. 
This originates from the fact that gravitational waveforms are not very 
sensitive to BH spins \citep{Abbott2017PhRvL}, unless the BH spins are high and precess around 
(are tilted with respect to) the binary angular momentum. In the four current 
BBH merger detections (GW150914, LVT151012, GW151226, GW170104) no 
significant precession was measured, and only in one event (GW151226) 
a non-zero spin magnitude of at least one BH is required. 

Our results show that accretion in progenitor binaries of massive BBH
mergers is not an efficient process to spin up BHs. For very low natal BH
spins ($a_{\rm in} \approx 0$), the final BH spin in not larger than 0.2--0.3, 
and for higher initial BH spins ($a_{\rm in} \gtrsim 0.5$) the BH spin 
remains almost unaffected by accretion. This is a direct result of the
fact the accretion onto a BH in the common envelope phase (the major accretion 
event for typical BBH progenitor) is not estimated to occur at rates 
much higher than a few percent of the Bondi rate \citep{Ricker2008, MacLeod2015, 
Murguia2017}. In other words, our results show that LIGO will be
measuring the natal BH spin in BBH mergers if they originate from the
classical formation scenario discussed here \citep{Belczynski16N}.
This is of fundamental importance, since BH natal spins can be
potentially used to infer information about the angular momentum/rotation of
the massive stars progenitors of these BHs. Rotation and angular momentum transport
in massive stars are basic missing blocks of stellar evolution theory. 

If, for example, LIGO BHs are found to typically have very low spins,
it will indicate that angular momentum is lost rather effectively from
massive stars, and the Tayler-Spruit dynamo \citep{Spruit1999} may be
a process that operates in massive stars. However, that would also
indicate that LIGO BHs originate from a different population than
Galactic and extra-galactic High-Mass X-ray Binaries, which all show
very high spins ($a\gtrsim0.9$; see Tab.1 in \citealt{Fragos2015}).
If, on the other hand, BH spins were to be found to be generally high,
it would mean that angular momentum transport is a rather inefficient
process (i.e. the core must decouple from the envelope early on in the
massive star evolution).  In this last case, a variety of long GRB
central engines based on single massive star progenitors (e.g.,
\citealt{MacFadyen1999}), could also be realized in BBH mergers.
Last, if a mixed population (low and high spins) of LIGO BHs is found
by future detections, it will point to mild coupling of layers within
a massive rotating star, moderated by the effects of wind mass loss
that depends on star mass and metallicity (e.g., Geneva stellar
models; \citealt{Eggenberger2012, Georgy2013}). Such a model would
naturally explain high spins of low mass BHs (like the ones observed
in HMXB in very local Universe), along with low spin BHs that have
high masses ($M_{\rm BH} \gtrsim 20-30 \msun$;
\citealt{Belczynski2017}).

 \begin{figure}
 \hspace{-0.3in}
\includegraphics[scale=0.44]{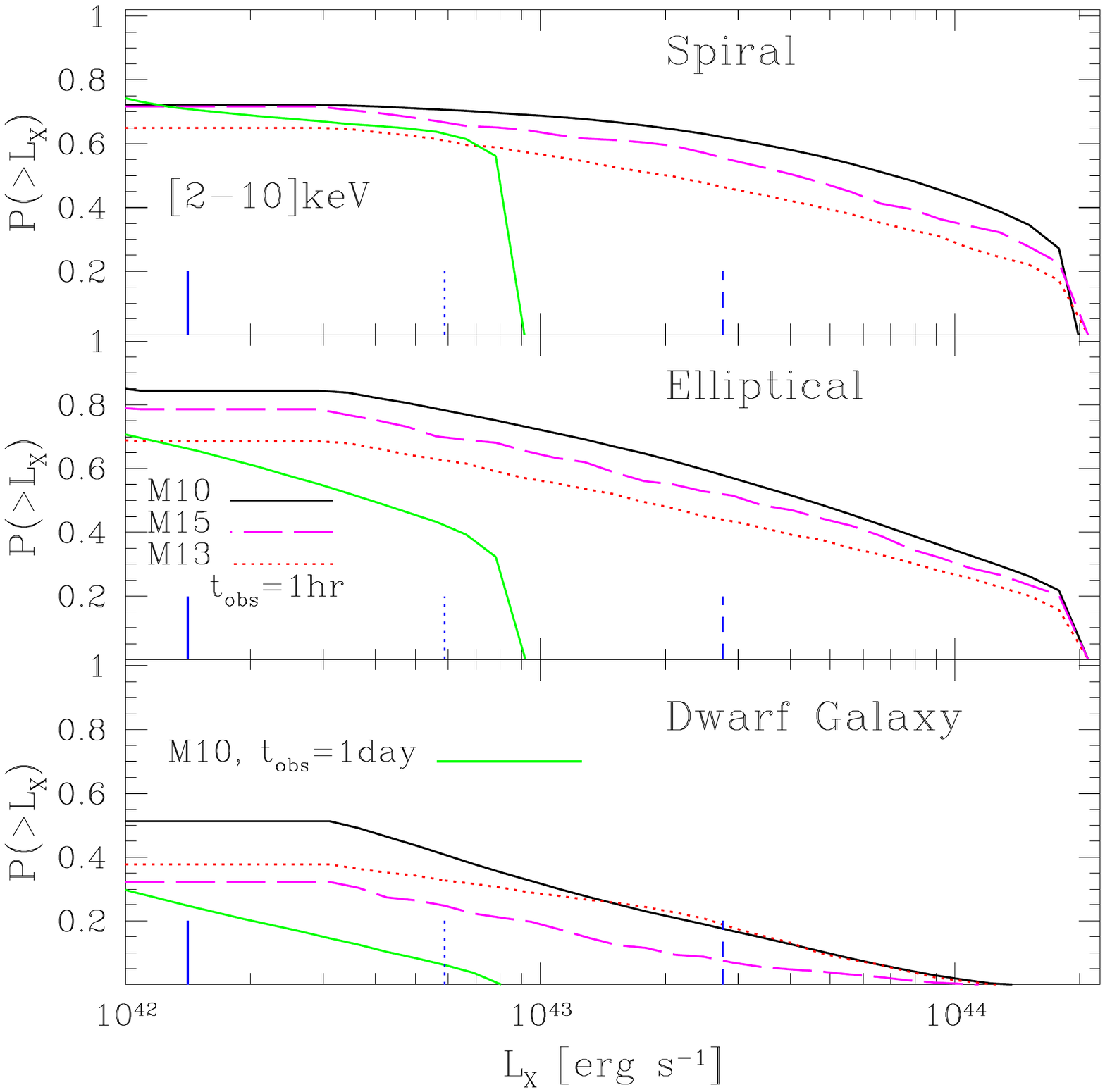}
\vspace{-0.5in}
\caption{The cumulative probability distribution of afterglow
  luminosities in the 2-10~keV band at 1~hr after the burst for our
  three representative models, { and at 1~day for model M10, as a
    reference for comparison to past observations}. The vertical lines
  indicate the minimum luminosity that would be detectable by
  \textit{Swift}/XRT (unabsorbed 2-10\,keV flux of $F_{\rm lim}\approx
  2.5\times 10^{-13}~{\rm erg}~{\rm s}^{-2}~{\rm cm}^{-2}$, see text)
  for a burst located at $z=0.05$ (solid line), $z=0.1$ (dotted line),
  $z=0.2$ (dashed line), respectively. Note that the probabilities do
  not saturate at 1, reflecting the fraction of much dimmer afterglows
  resulting from mergers outside of the host galaxy, in the
  intergalactic medium. This probability would need to be further reduced
  by the beaming factor if the emission is produced within a collimated jet. 
  {In particular, since the X-ray luminosity is expected to be produced 
when the flow is still relativistic,  those probabilities should be reduced by a factor of $\sim 3$,
with the beaming factor  constrained by the $\gamma$-ray limits on all the BBH GW merger events
  detected so far (cfr. \S4.1).}}
\label{fig:LX}
\end{figure}

 \begin{figure}
\includegraphics[scale=0.42]{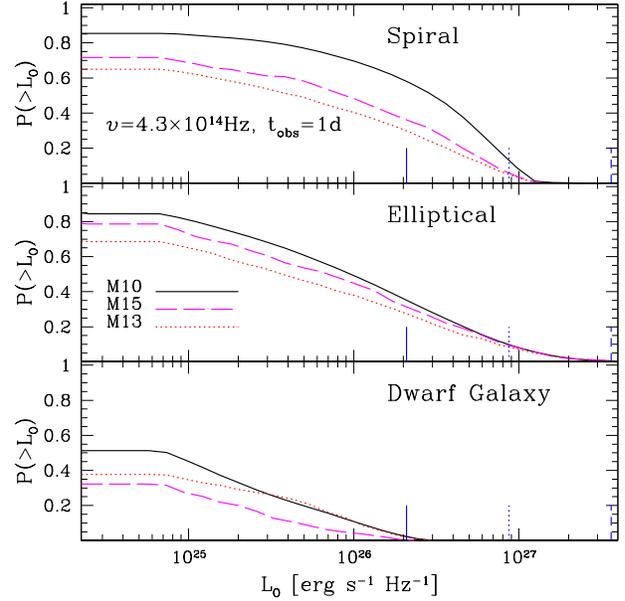}
\vspace{-1in}
\caption{Same as in Fig.\ref{fig:LX}, but at the frequency of
  $\nu=4.3\times 10^{14}$~Hz (falling in the R band) at 1~day after
  the burst. A limiting flux of $F_{\rm lim}\approx 4\,\mu$Jy
    (see text) has been assumed to set the values of the minimum
    luminosity for detection at a redshift of $0.05$ (solid line),
    $0.1$ (dotted line), $0.2$ (dashed line). {Some reduction in the probability
    due to beaming should be accounted for also in the optical, albeit at a lower level
    than in the X and $\gamma$-rays.}}
\label{fig:LO}
\end{figure}

 \begin{figure}
\includegraphics[scale=0.42]{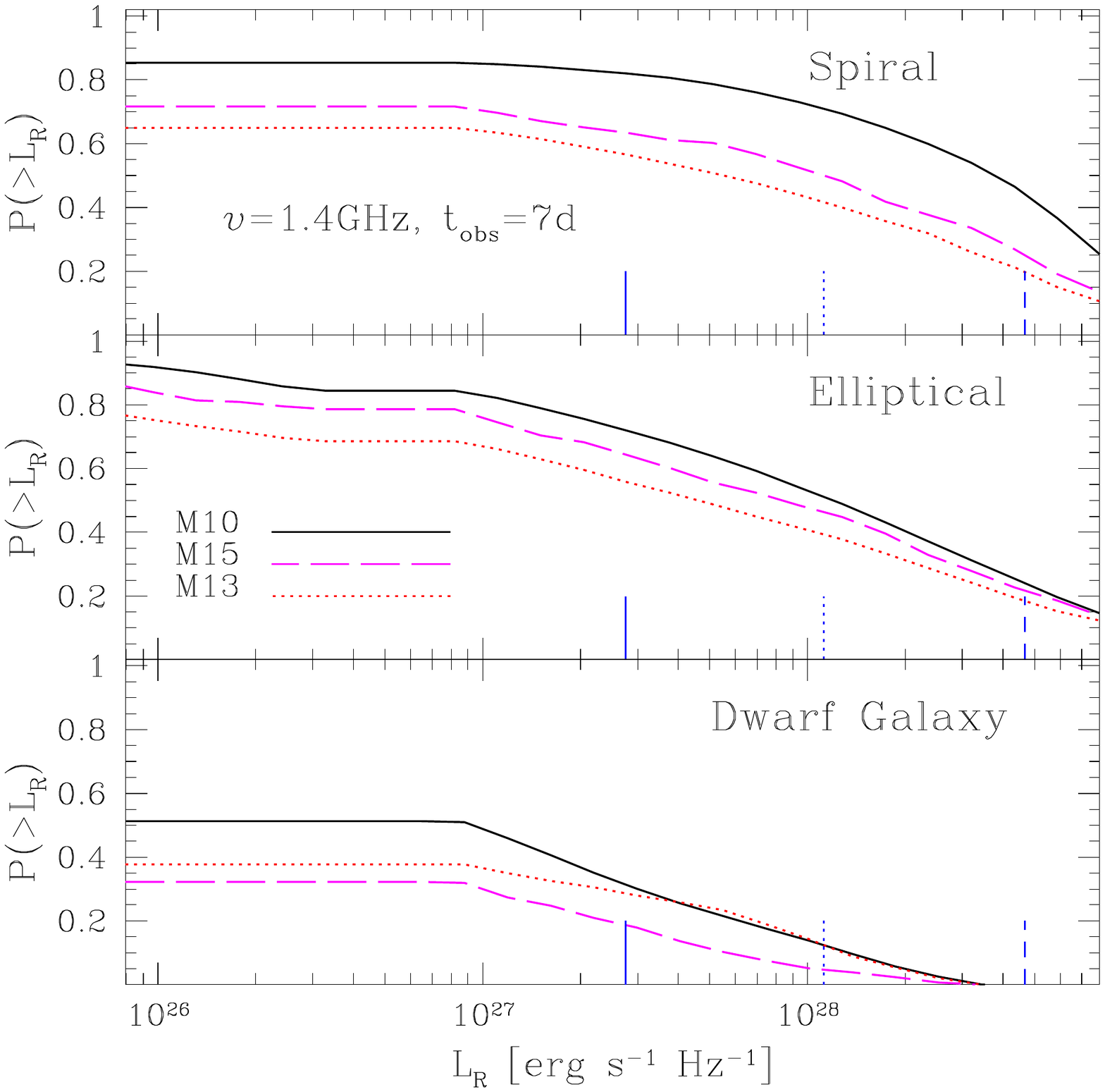}
\vspace{-1in}
\caption{Same as in Figs.~\ref{fig:LX} and \ref{fig:LO} but in the
  radio (1.4~GHz) at 7~days after the burst. A limiting flux of $F_{\rm
    lim}=50\mu$Jy (see text) has been assumed to set the
  values of the minimum luminosity for detection at a redshift of
  $0.05$ (solid line), $0.1$ (dotted line), $0.2$ (dashed line). {In the radio band,
  the emission is expected to be close to isotropic, and hence the computed probabilities
  are not expected to be affected by beaming effects.} } 
\label{fig:LR}
\end{figure}

\subsection{EM counterparts to BBH mergers}
\label{subsec:results}
\subsubsection{Predictions for energetics comparable to that of the \textit{Fermi} counterpart}

In the following, we use the radiation model described in Sec.\,\ref{subsec:model}, coupled with the
density distribution at the BBH merger sites calculated in
Sec.~\ref{subsec:BinEv}, to estimate the expected brightness
distribution of afterglows possibly associated with BBH merger
events. 

We compute the predicted afterglow luminosities under the assumption
that each BBH merger event is associated with an explosive release
of energy. For simplicity, {and for the lack of other reference numbers}, we take as a representative case
the energetics of the candidate $\gamma$-ray counterpart to GW150914
reported by the \textit{Fermi} collaboration \citep{Connaughton2016},
i.e. $E=10^{49}$~erg. { We remind the reader  (cfr. Sec.4.1) that this implies 
that the high energy $\gamma$-ray emission must be produced within a jet of 
angular size $\lesssim 50^\circ$ and be relativistic beamed.}
We  adopt $\xi_B=0.01$ and $\xi_e=0.1$, as
typical values inferred from broad-band afterglow modeling in the
'standard' GRBs \citep{Wijers1999,Panaitescu2001}. 
Using a single value for $E,\xi_B,\xi_e$ (rather
than assuming some arbitrary distributions) allows us to \textit{(a)}
highlight the dependence on the different evolutionary models (which
predict different merger sites and hence medium densities), and
\textit{(b)} present results that can be immediately rescaled to other
values of these parameters (which may be motivated by future
observations), thanks to the analytical dependence of the afterglow flux
on these parameters (see Sec.4).

Fig.~\ref{fig:LX} shows the distribution of afterglow luminosities in
the 2-10~keV band for our three
representative models and three galaxy types. Since, as discussed
above, the merger sites have very little dependence on metallicity, we
only show results for $Z=10\%Z_\odot$.  
{In addition to showing predictions at an observing time of 1~hr
after the merger, as would be typical following a $\gamma$-ray localization,
we also display, for the reference M10 model, the expected flux distribution
at 1~day after the merger, to reflect what realized with the X-ray follow ups
to the LIGO-detected events in runs O1 and O2.}
The vertical lines indicate
the minimum luminosity that would be detectable by \textit{Swift}/XRT
in a follow-up search with sensitivity similar to the one achieved for
GW150914 \citep[$F_{\rm lim}\approx 2.5\times 10^{-13}~{\rm
    erg}~{\rm s}^{-2}~{\rm cm}^{-2}$ in the 2-10\,keV band;
  see][]{Abbott2016c} for a burst located at the three representative
redshifts of $z=0.05, 0.1, 0.2$ (all within the LIGO BBH horizon). The
probability of detecting an X-ray counterpart is in the range of
65-85\% depending on the evolutionary model and galaxy type for the
closest elliptical/spiral galaxy, and between 40-70\% for the farthest
galaxy. A larger fraction of brighter events is expected in the M10
model, due to the generally smaller distances (and hence higher
densities) of the merger sites from the galaxy centers. Note that the
probabilities flatten out at low luminosities with values $<1$ as a
result of fact that a fraction of BBH mergers is predicted to occur
outside of the host galaxy, in the intergalactic medium.  The
luminosities of those events (which would bring the total cumulative
probabilities to 1) are lower than the lowest limit in the figure. In
the case of dwarf hosts, for the same emission parameters ($E,\xi_e,
\xi_B$) used for the massive galaxies, the luminosities are lower, due
to the smaller number densities. Additionally, a larger fraction of
events is expected to be 'naked', due to the mergers occurring outside
the host. This fraction can be directly read from the flattening of
the probability distributions at low luminosities: up to a fraction
$\sim$ 30-50\% (depending on the magnitude of the natal kicks and the
redshift of the host) could have a non-negligible afterglow-like EM
counterpart.

Fig.~\ref{fig:LO} shows the same distributions, but in the optical
after 1~day. The vertical lines, corresponding to the minimum
detectable luminosity for events at the same representative redshifts
as in Fig.~\ref{fig:LX}, have been computed assuming a $5\sigma$
limiting magnitude of $r\approx 22.4$\,mag (AB system, equivalent to a
limiting flux of $F_{\rm lim}\approx 4\,\mu$Jy), comparable to the
most sensitive optical searches carried out in this band during the
follow-up of GW150914 \citep[see Table 1 in][]{Abbott2016c}.  An
inspection of Fig.~\ref{fig:LO} shows that detection in the optical is
not likely for a merger in a dwarf galaxy, even at a low redshift of
$z\sim 0.05$.  For massive galaxies, the detection probability becomes
more sizable, but only at low redshifts. In particular, for $z=0.05$,
a fraction between 30-60\% of events could be detected in the optical.
This fraction becomes 10\% or smaller already at $z=0.1$.

Fig.~\ref{fig:LR} shows the probability distributions in the radio
band, at 1.4~GHz after 1 week from the explosive event. Here,
  the reference threshold flux corresponding to the minimum detectable
  luminosities at the redshifts of 0.05, 0.1, 0.2 is set at the
  $5\sigma$ threshold flux for the Karl G. Jansky VLA (A
  configuration) for 1~hr integration on-source ($F_{lim}\approx 50\,\mu$Jy).
For mergers at a redshift of 0.2, corresponding to the highest
redshift LIGO BBH merger so far \citep{Abbott2017}, about 25-45\% of events
  could be detected as radio afterglows, depending on the model and
  galaxy type, for massive galaxies, while for dwarf galaxies the detection
probability is non-negligile only at lower redshifts, being about 20\% for events
at $z=0.1$.  
Note that, for the particular choice of afterglow parameters adopted
here (namely $E, \xi_e, \xi_B$), and the range of densities probed by
the merger events, the afterglow spectrum is in the regime described
by the middle branch of Eq.~(\ref{eq:Fnu2}). In this case, the flux at
the same observation time is higher at lower frequencies. However,
given the sensitive parameter dependences of Eq.~(\ref{eq:nuc}) and
Eq.~(\ref{eq:num}), the frequency-dependence across the bands can be
reversed for other combinations of the relevant parameters.

We finally note that, up to this point, our calculations have estimated the brightness of
possible afterglow counterparts independently of their visibility due
to beaming effects.  However, an afterglow-like EM counterpart to a GW
event may come from a relativistic jet
of size $\theta_{\rm jet}<\pi/2$, {and the discussion of \S4.1 has shown that,
for an emission model with energetics comparable to those inferred for GW150914,  
the high energy emission must be confined within a jet of $\lesssim 50^\circ$. This
constraint, which was derived for the $\gamma$-ray emission from the \textit{Fermi}
 follow ups during the O1/O2 runs,
likely extends also to the
X-rays, which are produced when the shock  is still relativistic (and more moderately to
the optical). } Therefore 
our computed probabilities for the observability of afterglow-like
emission should be  reduced by the beaming factor $(1-\cos\theta_{\rm jet})$ { $\sim 1/3$ in the
X-rays}  if
interpreted as detection probability of an EM counterpart
following a GW event independently of a $\gamma$-ray detection. 
{A milder reduction is expected in the optical, while the radio emission should be
close to isotropic.}

\subsubsection{Model constraints from past observations}
Among the LIGO BBH events, GW150914 was probably the one that enjoyed
the most extensive and coordinated EM follow-up observations
\citep{Abbott2016c}, and the one for which a possible association with
a gamma-ray transient was also claimed \citep{Connaughton2016}. Here
we make the assumption that the EM follow-up effort that accompanied
GW150914 is representative of the typical effort that will be put in
place by the community in the future hunt for EM counterparts to BBH
mergers. In this framework, it is thus reasonable to assume that, as
for the case of GW150914, optical surveys will be able to cover
typical localization areas of order $\sim 100$\,deg$^2$ down to
$r<22.4$\,mag within a timescale of order $\sim 1$\,d since the merger
\citep{Abbott2016c}.  X-ray observatories such as e.g. \textit{Swift}
may instead cover localization areas of $\lesssim 5$\,deg$^2$ within a
timescale of $\sim 1$\,day since the merger and down to a limiting
2-10 keV unabsorbed flux of
$2.5\times10^{-13}$\,erg\,cm$^{-2}$\,s$^{-1}$ \citep{Abbott2016c}.

For what concerns typical GW localization areas, for massive BBHs
located at $z\lesssim 0.1$, and as demonstrated in the case of
GW170814 \citep{Abbott2017a} we can expect the two LIGO and Virgo
detectors to provide 90\% credible areas of $\sim 60$\,deg$^2$. Thus,
while optical follow-up observations will be able to cover basically
all of of the GW localization areas, deep X-ray follow-up observations
may only end up covering $\lesssim 10\%$ of the GW localization area.

Based on the above considerations and on our estimates, optical
follow-ups of massive BBHs at $z\lesssim 0.1$ will be limited by
the theoretical expectations that only $\approx 5-10\%$ of the BBH
mergers located in spiral/elliptical galaxies are expected to have an
optically detectable ($r<22.4$\,mag) afterglow. Thus, in order to
exclude the predictions presented in this work at the $\gtrsim 90\%$
confidence, we would need to collect a sample of $\sim 25-50$
 BBH events {i.e. with energetics comparable  } GW150914
with deep optical follow-up observations of
their GW localization areas. Our estimates also show that a fraction
$\approx 40-70\%$ of the BBH mergers in spiral/elliptical galaxies at
$z\lesssim 0.1$ would have an X-ray afterglow detectable at 1\,d since
merger. But, for these we expect only $\sim 10\%$ of the localization
area to be covered with X-ray follow-up observations. Thus, overall,
we would need a sample of $\sim 30-60$ GW150914-like BBH events with
deep X-ray follow-up observations to exclude the predictions presented
here at the $90\%$ confidence level. {Applying the same constraints
on beaming as derived from the $\gamma$-rays, the number of BBH events 
quoted above would need to be multiplied by a factor of about 3.}

In summary, because it appears likely that a sample $\sim 50-100$ BBH
events will be collected by the advanced detectors in their
upcoming observing runs \citep{Abbott2016d},
testing the predictions presented here is in fact within reach, as
long as optical/X-ray follow-up efforts continue to be pursued by the
community. The above considerations also demonstrate that so far the
lack of detections of radio/optical/X-ray afterglows from the 4 significant
LIGO/Virgo BBH detections (e.g. \citealt{Copperwheat2016,Evans2016,Kasliwal2016,
  Morokuma2016,Palliyaguru2016,
  Smartt2016a,Smartt2016b,Bhalerao2017,Corsi2017,Kawai2017}) does not rule out the hypothesis
that some or all of GW150914-like events may be accompanied by
gamma-ray transients giving rise to multi-band afterglows.

Finally, we note that while our estimates for the detectability of
radio afterglows with sensitive GHz telescopes such as the VLA are
promising ($\gtrsim 30\%$ for BBHs located in spiral/ellipticals at
$z\lesssim 0.1$), these telescopes also have small field of views and
thus the EM follow-up at these wavelengths will likely rely on
accurate localizations provided by the identification of optical or
X-ray counterparts.

\section{Summary}

In light of the detection of GWs from BBH mergers, and the massive
efforts at searching for EM counterparts from such events, here we
have performed a study aimed at predicting the statistical properties
of the merged BBH population within their host galaxies, for BHs
produced via massive stars evolution in isolated binaries.  
In addition to calculating the mass function (provided in terms of both chirp
and total mass), for three values of the metallicity, we have computed the spin
distribution of the primary BH as a result of accretion of mass from
the secondary during the close binary evolution.  Of particular
importance for optimizing follow-up strategies, we provided the
expected distance distribution within the galaxy potential,
for a massive spiral, a massive elliptical, and a dwarf galaxy
  (modeled as a scaled-down version of a spiral).  We considered the
two most extreme models (as well as one in between) for the kicks
received by the BHs at birth. As such, our results span the expected
range for BBH mergers within the isolated binary scenario. 
The distribution of the merger sites further allows us to compute (for a given
galaxy model) the ISM density distribution for the merger locations.

{Last, we entertained the possibility that BBHs may be accompanied
by the formation of an outflow and a release of electromagnetic
energy via a shock. This was motivated by the tentative detection
of a $\gamma$-ray counterpart to GW150917, and a number of ideas
and models aimed at explaining such emission (see \S1 and \S4.1 for references
and details).
However, the lack of detection of $\gamma$-ray counterparts to the
other 5 BBH GW events (including the LIGO/Virgo trigger) implies 
that either the $\gamma$-ray luminosities are lower, or that
the emission is beamed within an angle $\lesssim 50^\circ$.  
By adopting (for simplicity) the $\gamma$-ray luminosity inferred for the
GW150917 candidate counterpart, we have hence made the
implicit assumption that the $\gamma$-ray emission is beamed to within $\lesssim 50^\circ$. 
With this caveat in mind, and the computed distributions of ISM densities at the merger locations,
we could then calculate the broad-band afterglow luminosities expected from these events.
Our estimates are particularly instructive since, especially at the longer (radio) wavelength,
beaming is less important and hence observational limits can help set tighter constraints
on proposed emission models. }

Our results can be summarized as
follows:
\begin{enumerate}
\item[\textit{(i)}] The mass function has a strong metallicity
  dependence, as expected.  A chirp mass $\sim 20~M_\odot$ requires
  metallicities $\lesssim 0.1~Z_\odot$.  The influence of natal kicks
  is much weaker than the influence of metallicity on the BBH merger
  mass function.
\item[\textit{(ii)}] The accreted mass from the progenitor star of the
  secondary BH to the primary BH during the period of mass transfer
increases the BH spin by an amount which, in the standard model M10 
is about 0.1 for 0.01\% and 0.1\% $Z_\odot$, and about 0.2 for 
$Z_\odot$.  BHs formed at high metallicity are less massive, and hence easier
to spin up for the same amount of accreted mass.
Additionally, the spin increase, for the same amount of accreted mass, is slightly
smaller for larger initial natal spins of the primary BH (see also \citealt{Belczynski2017}). 
For the same initial conditions, models M13 and M15 produce a generally
larger fraction of high spin BHs, with a flat spin distribution
up to $a\sim 0.35$ if the initial spin of the primary BH is zero.
\item[\textit{(iii)}] The distribution of merger sites (and hence of
  projected distances/angular offsets) within the host galaxies spans
  a large range of values, with a fraction $\sim 10-30\%$ for massive
  galaxies (with precise value depending on the model and galaxy type)
  occurring outside of the galaxy, at distances $R\gtrsim 100$~kpc.
  This fraction of mergers outside of the hosts is larger ($\sim
  40-60\%$) for a dwarf galaxy, here modeled as a spiral with mass of
  0.1\% that of the Milky Way.  As expected, in the model with the
  largest natal kicks (M13), there is a larger fraction of mergers
  occurring outside of the galaxy, and more generally at larger
  distances from the center. These offsets could be easily
    measured by any observational facility with sub-arcsec
    localization capabilities, if the merger is associated with EM
    emission.
\item[\textit{(iv)}] 
Using the distribution of interstellar medium densities derived from
the distribution of merger sites within the host galaxies, and
assuming that BBH merger events produce GRB-like counterparts similar to
the \textit{Fermi} candidate counterpart to GW150914, we have
computed the resulting afterglow distributions in the X-rays, optical,
and radio band.  We find that, for massive galaxies, a fraction on the
order of $10-30\%$ of the mergers (where the precise value depends on
the assumed natal kick distribution and galaxy type) is expected to be
naked, i.e. lacking a bright afterglow at any wavelength. The same
fraction is higher, $\sim 40-60\%$, in dwarf galaxies.  These events
correspond to mergers that occur outside of the host galaxy, within
the diffuse intergalactic medium.

A fraction of the order of $\sim 40-70\%$ (depending on the model and
galaxy type and before beaming corrections) of the mergers in 
massive galaxies would be accompanied by broad-band
afterglow-like emission detectable in X-rays if occurring within the
LIGO horizon and if associated with a $\gamma$-ray signal similar to
the \textit{Fermi} candidate counterpart to GW150914 (which would require a kinetic
energy on the order of $10^{49}$~ergs). A smaller
fraction could also be detectable at radio and optical wavelengths
with sensitive enough follow-up instruments. 
 Given the analytical
scaling of afterglow flux with energy, our results can be easily
scaled should future observations indicate positive detection but
point to different values of the energetics.

{Our goal for these calculations has been that of providing a
  theoretical framework against which to test ideas for impulsive
  energy production accompanying BBH mergers, given the expectation
  that a sudden energy release would be accompanied by an afterglow-type radiation, in
  analogy to what observed in connection with the binary neutron star
  merger event \citep{Abbott2017a}. To this purpose, we specifically convolved our
  theoretical predictions with the observational set up (i.e. time of
  the observations, sensitivity, sky coverage) of past EM follow ups
  to LIGO detections of BBH mergers, and assuming this to be typical
  of future observational campaigns, we provided an estimate of the
  number of BBH mergers that needs to be followed in order to rule out
  explosive-like events accompanying BBH mergers. In particular, we
  found that we would need to collect a sample of $\sim 25 - 50$
  GW150914-like BBH events with deep optical follow-up observations of
  their GW localization areas, and $ \sim 30- 60$ GW150914-like BBH
  events with deep X-ray follow- up observations, to exclude the
  predictions presented here at the 90\% confidence level.}

\end{enumerate}

{\bf Acknowledgments}.
{We thank an anonymous referee for very useful comments and suggestions, 
and Valerie Connaughton
for providing us with the details of the \textit{Fermi} GBM follow ups to the BBH merger events
in runs O1 and O2.}
RP was partly supported by NSF award AST-1616157.
KB acknowledges support from the Polish National Science Center
(NCN) grant: Sonata Bis 2 (DEC-2012/07/E/ST9/01360). AC acknowledges support from the NSF CAREER award \#1455090
and also partial support from the Swift Cycle 12 GI program (Grant No. NNX17AF93G).


\end{document}